\documentclass[superscriptaddress,groupedaddress,nofootnoteinbib,12pt]{article} 
\pdfoutput=1
\usepackage{color}
\usepackage{colortbl}
\usepackage[pdftex]{graphicx}  
\usepackage{dcolumn}   
\usepackage{bm}       
\usepackage{amssymb}  
\usepackage{amsmath}
\usepackage{sectsty}
\usepackage{latexsym}
\usepackage{float}
\usepackage{ifthen}
\usepackage{caption,subfig}
\usepackage{enumerate}
\usepackage{url}
\usepackage{caption,subfig}
\usepackage{feynmf}	
\usepackage{jcappub}
\usepackage{amsopn}
\usepackage{float}
\usepackage{pst-pdf}
\usepackage{amsfonts}
\usepackage{multirow}
\usepackage{array}
\usepackage{booktabs}
\usepackage{rotating}

\def\clap#1{\hbox to 0pt{\hss#1\hss}}

\def\({\left(}
\def\){\right)}
\def\[{\left[}
\def\]{\right]}
\def\bea{\begin{eqnarray}}
\def\eea{\end{eqnarray}}
\def\be{\begin{equation}}
\def\ee{\end{equation}}
\def\ba{\begin{eqnarray}}
\def\ea{\end{eqnarray}}
\def\beq{\begin{eqnarray}}
\def\eeq{\end{eqnarray}}
\def\mpl{M_{\rm Pl}}
\def\d{\mathrm{d}}

\def\clap#1{\hbox to 0pt{\hss#1\hss}}

\newcommand{\etal}{et al.}
\newcommand{\fNL}{f_{\mathrm{NL}}}

\definecolor{forestgreen}{rgb}{0.133,0.545,0.133}

\renewcommand{\d}{\mathrm{d}}
\renewcommand{\vec}[1]{\bm{\mathrm{{#1}}}}

\newcommand{\para}[1]{\par\vspace{2mm}\noindent\emph{{#1}}.---}

\newcolumntype{Q}{>{$\displaystyle}l<{$}}
\newcolumntype{q}{>{\columncolor[gray]{0.9}$\displaystyle}l<{$}}
\newcolumntype{R}{>{$\displaystyle}r<{$}}
\newcolumntype{S}{>{$\displaystyle}c<{$}}
\newcolumntype{s}{>{\columncolor[gray]{0.9}$\displaystyle}c<{$}}
\newcolumntype{T}{>{\columncolor[gray]{0.9}}c<{}}

\newsavebox{\tableA}
\newsavebox{\tableB}
\newlength{\tblw}

\newsavebox{\boxplot}
\newsavebox{\boxplota}

\newcommand{\tablepreamble}{\small%
	\heavyrulewidth=.08em%
	\lightrulewidth=.05em%
	\cmidrulewidth=.03em%
	\belowrulesep=.65ex%
	\belowbottomsep=0pt%
	\aboverulesep=.4ex%
	\abovetopsep=0pt%
	\cmidrulesep=\doublerulesep%
	\cmidrulekern=.5em%
	\defaultaddspace=.5em%
	\renewcommand{\arraystretch}{1.6}}

\newboolean{editorial}
\setboolean{editorial}{true}
\newcommand{\editorial}[2]{\ifthenelse{\boolean{editorial}}{\textcolor{dullpurple}{[\textsf{\textbf{{#1}}}: }\textcolor{blue}{\textsf{{#2}}}\textcolor{red}{]}}{}}

 \def\be   {\begin{equation}}   \def\ee   {\end{equation}}

 \def\ba  {\begin{eqnarray}}   \def\ea  {\end{eqnarray}}

\renewcommand{\thesubsection}{\arabic{section}.\arabic{subsection}}

\newcommand{\eq}{\end{equation}}
\newcommand{\bi}{\begin{itemize}}
\newcommand{\ei}{\end{itemize}}

\def\d{{\rm d}}

\def\imath{{\rm i}}

\begin{document}

\title{The effects of a fast-turning trajectory \\ in multiple-field inflation
}

\author{Maciej Konieczka,$^{a}$ Raquel H.~Ribeiro,$^{b}$ and Krzysztof Turzy\'nski$^{a}$}

\affiliation{$^{a}$Institute of Theoretical Physics, Faculty of Physics, University of Warsaw, \\
ul. Pasteura 5, 02-093 Warsaw, Poland}
\affiliation{$^{b}$CERCA/Department of Physics, Case Western Reserve University, \\ 10900 Euclid Ave, Cleveland, OH 44106, USA}
    \emailAdd{maciejkonieczka@student.uw.edu.pl}
	\emailAdd{RaquelHRibeiro@case.edu}
    \emailAdd{Krzysztof-Jan.Turzynski@fuw.edu.pl}

\abstract{
The latest results from  
\emph{Planck} 
impose strong constraints on features in the
spectrum of the curvature perturbations from
inflation.
We analyse the possibility of particle production induced by sharp turns 
of the trajectory in  
field space
in 
inflation models with multiple fields.
Although the evolution of the background fields can be altered by
particle production,
we find rather modest changes in the power spectrum even for
the most extreme case in which the entire kinetic energy
of the scalar fields is converted into particles.
}

\maketitle


\section{Introduction}
\label{sec:intro}
After an era characterised by scarcity and inaccuracy of data, cosmology is 
now thriving with observations and measurements, which make our understanding 
of the universe remarkably precise. A key feature of our present picture of the early universe 
is the presence of small inhomogeneities 
of energy density.  
Whilst their exact origin cannot be directly inferred from the data, it certainly lies within 
the quantum realm. An almost universally accepted opinion is that they appeared as quantum 
fluctuations of one or more scalar fields and were subsequently amplified during a period of inflation
to become seeds for the anisotropy 
pattern in the microwave sky and the large structures in the universe \cite{Guth:1980zm,Linde:1981mu,Mukhanov:1981xt}.

The latest and 
most accurate data, obtained by the \emph{Planck} collaboration
\cite{Ade:2013ydc,Ade:2013zuv,Ade:2013uln}, 
strongly support the paradigm of inflation, 
which generically predicts 
a non-scale-invariance of the spectrum of the primordial perturbations. 
Present observations rule out
a scale-invariant spectrum 
at a significance level exceeding 5 standard deviations. 
In spite of this progress, the data suggest a large degeneracy between predictions of different inflationary models. 
This should be, to some extent, expected 
as the power spectrum represents the lowest order statistics, and
necessarily loses a vast amount of information (in particular, about the interactions 
of these fields). Some exceptions         
                                        exist, namely in 
                                        models of inflation
                                        in which the effects of 
                                        a heavy scalar field
                                        can be accommodated by a non-trivial
                                        speed of sound of perturbations
                                        for a massless one (see, 
                                        for example, 
                                        Refs. \cite{Cremonini:2010ua,
                                        Achucarro:2010jv, 
                                        Achucarro:2010da,
                                        Avgoustidis:2012yc}).

Whereas this might indicate that 
single-field models can be acceptable toy models for inflationary phenomenology, on a theoretical 
level these models lack a well-defined setup. 
This is because in a single-field model
there is a lack of control of the
quantum-mechanically induced radiative corrections, 
which can change the shape of the potential, 
thereby changing predictions for observables. 
This issue
can be ameliorated if we take an effectively single-field action
as arising after integrating out heavy degrees of freedom, 
which couple to the light field.
Moreover,
in high-energy models of particle physics one expects a number of light scalar degrees of freedom, 
with masses not much larger (and often smaller) than the Hubble scale. The fluctuations of these fields 
can, in principle, act as sources of primordial perturbations. This feature has led to a significant investment 
in searching for characteristic signatures in models with multiple light scalar fields. One of the key aspects of 
such models is that, unlike in single-field inflation scenarios, the primordial curvature perturbation need not be 
frozen on super-horizon scales \cite{Bardeen:1983qw,Kofman:1986wm}, 
which makes the phenomenology of such models potentially richer and more interesting.

There are  
a number of possible effects, which result from the presence of multiple active scalar 
degrees of freedom during inflation. One well-studied example is the coupling between the fields, 
either non-minimal or through the metric itself, which might cause the inflationary trajectory to deviate from 
a geodesic line in  
field configuration space. Such bending can then induce 
a change in the power spectrum of perturbations, which ought to be properly taken into account for a rigorous 
comparison to observations \cite{Chen:2009zp, Peterson:2010np, Achucarro:2010da}. 
In addition, some of the usual slow-roll parameters can become large, which requires 
keeping track of higher-order 
corrections \cite{Peterson:2010np,Avgoustidis:2011em}. 

On the other hand, the coupling between different fields can also give rise 
to observable non-gaussianity  
with quite a 
distinctive statistical signature, making it possible to test these 
models, especially with the
current stringent constraints  \cite{Ade:2013ydc}. 
Nevertheless, it is fair to argue it will be  
possible to choose a region in the 
parameter space for which the predicted non-gaussianity is well within the observational constraints. This possibility 
makes the degeneracy between different inflationary models much broader and extends it beyond 
the two-point statistics \cite{Elliston:2013afa}.

Even in the absence of directly observable effects, 
understanding the possibly rich phenomenology in multiple-field inflation 
can also provide an important toolkit for grasping the detailed physical picture 
of the inflationary mechanism. 
This is somewhat analogous to the 
relatively recent motivation behind analysing theories of modified gravity in the IR
(see, e.g., Clifton \etal~\cite{Clifton:2011jh} for a review). 
Whilst the theory of general relativity agrees very well with a vast array of observations, 
studying various modifications of this theory allows gaining more insight into theories of gravity 
as effective field theories.\footnote{The currently observed
                                    accelerated expansion of 
                                    the universe
                                    can be an indication of 
                                    GR breaking down in the IR, 
                                    which adds to the interest
                                    in this research area.
                                    In the inflationary cosmology
                                    case, 
                                    on the other hand, 
                                    there is still no 
                                    empirical hint for a breakdown
                                    of the single-clock picture.}

For these reasons
it is important to 
investigate the potentially interesting physical effects
 arising 
in these models. 
In this regard, there is a noticeable gap in the existing literature on the subject. 
Whenever the evolution of the background fields is non-adiabatic, particle production can occur
(see, e.g., Refs.\cite{Watson:2004aq,Kofman:2004yc,Cremonini:2006sx,Green:2009ds,Battefeld:2011yj,Enomoto:2013mla}
for interesting realizations of this concept); 
this effect may drain the energy from the scalar field 
sector and transform it to a component of the total energy density with a different evolution. 
The non-adiabaticity can manifest itself in fast changes of the masses of the field(s)
or, in multiple-field models, in a fast change of the direction 
of the mass eigenstates in  
field space.
Strikingly, 
despite the fact that relevant theoretical tools are readily available
\cite{Nilles:2001fg},
only the first of the two possibilities has been examined in the context of inflation, 
in setups known as moduli trapping scenarios.  
Filling this gap will constitute a step towards a better 
understanding of the phenomenology in multiple-field models of inflation.

In this paper, we shall therefore analyse 
the consequences of particle production in the simplest multiple-field models of inflation 
in which the inflationary trajectory exhibits a fast turn.
In particular, transferring energy into particles during inflation can affect the 
homogeneous background, thereby changing the inflationary trajectory as well as the evolution of the parameters, 
such as
the Hubble parameter.
Consequently,
the predictions of many important cosmological observables 
(power spectrum of primordial perturbations, normalization and shape of potential non-gaussianities) 
may be modified. Obtaining such accurate predictions with particle production fully taken into account 
will allow for a better insight into the phenomenology of multiple-field models of inflation. 
The primary purpose of this paper is to provide an advance in this direction.

\para{Outline}This paper is organised as follows. In \S \ref{sec:f1} we
analyse the behaviour of perturbations in the vicinity 
of a trajectory turn in field configuration space, whilst 
ignoring the possibility of particle production. 
We drop this simplifying assumption in \S \ref{sec:pp}, 
and use the mode mixing technique to compute the 
Bogolyubov coefficients as a means to determine the 
efficiency of particle production. We provide both analytical
and numerical estimates by investigating a particular 
two-field inflation model
in which a sharp turn of the inflationary flow occurs.
We also study the backreaction of particle production on the classical 
inflationary trajectory.
In \S \ref{sec:particles} we elaborate on the interpretation of 
the occupation number in particles on the basis of the Bogolyubov
coefficients, and review the reasons why the concept of 
particle production in a time-dependent setting is unclear
from the field theory viewpoint.
We summarize and discuss the 
implications of our results in \S \ref{sec:discuss}.

\para{Notation}We use units in which the Planck mass, 
$M_\textrm{P}^2=1/8\pi G$, 
is set to unity, 
and $c=\hbar=1$. 
We work with the mostly plus metric signature 
and take the cosmological background to be given 
by the Friedmann--Lema\^{i}tre--Robertson--Walker line element. 
We use primes to denote 
derivatives with respect to conformal time, $\tau=\int a(t)\, \d t$, where $a$ is the scale factor, 
while overdots denote
differentiation with respect to cosmic time, $t$.
Lorentz indices are contracted using the metric 
$g_{\mu\nu}$ under Einstein's convention and 
in which Greek indices denote 
spacetime coordinates, whilst Roman letters specify the 
spatial components only. Capital Roman letters are reserved
to specify the particle species.
We adopt the usual
                            terminology that the dimensionless
                            power spectrum, $\mathcal{P}$, 
                            can be obtained from 
                            the traditional power spectrum, $P$, 
                            by applying the rule
                            $\mathcal{P}=k^3 P/2\pi^2$.


\section{Sharp turns without accounting for particle production}
\label{sec:f1}

Consider a model with multiple, canonically normalised 
scalar fields described by the action
\begin{equation}
S=\frac{1}{2} \int{\d^4 x \, \sqrt{-g}}\  
\Big\{R-(\partial_{\alpha}\phi)^2 -(\partial_{\alpha}\chi)^2 -2V(\phi, \chi)
\Big\}\ .
\end{equation}
The classical inflationary
trajectory in this case can be decomposed into infinitesimal 
displacements of each of these fields in configuration space, 
$\big\{\delta \phi, \delta \chi \big\}$. 
This trajectory will not be, in general, a straight line
since the fields will attempt to minimise the potential. 
This can be intuitively understood by describing the inflationary
flow by a bundle of geometrical rays which
respond to Huygens equation \cite{Seery:2012vj}---a light ray bends owing
to a varying refractive index, which can be shown to be related to the shape
of the potential 
(in particular, to the slow-roll parameter 
$\varepsilon\equiv -\dot{H}/H^2$).
  
How are the predictions for the power spectrum of the curvature
perturbations modified after a sharp turn of the inflationary trajectory?
In what follows we shall refer to a sharp turn if the 
rate of turning of the trajectory is
much larger than the Hubble parameter, so that the turn
completes in a fraction of an e-fold. Said differently, the
rate of turn is the largest
 scale entering the equations
of motion of the inflationary perturbations 
(or it is at least of the same order as the largest mass parameter).

The effects of 
such a sharp turn of the inflationary trajectory have been addressed by a number of authors 
(see, for example, 
Refs. \cite{Chen:2009zp, Achucarro:2010da,Cespedes:2012hu}) .
Interestingly, a couple of analyses found a correction to the power spectrum
which does not vanish in the limit of infinitely large  comoving
wavenumber $k$. 
In Ref.~\cite{Shiu:2011qw}, Shiu \& Xu have argued
that the correction to the 
dimensionless power spectrum
can be approximated as 
\begin{equation}
\mathcal{P}_{\zeta}/\mathcal{P}_0\simeq 1+2\Delta\theta \sin(2k/k_\mathrm{turn})\ \ ,
\end{equation}
where $\mathcal{P}_0=H^4_\ast/8\pi^2\varepsilon_\star^2$
is the (dimensionless) power spectrum in an effectively single-field 
inflation model. Here, 
$H_\ast$ and 
$\varepsilon_\star$ are the Hubble  
and the 
slow-roll parameters evaluated  
at 
Hubble  
crossing  
for the reference mode, 
which is taken to be the mode that leaves
the Hubble radius exactly at the turn; 
$\Delta \theta$ is the deflection angle of
 the inflationary trajectory.

In contrast, Gao, Langlois \& Mizuno \cite{Gao:2012uq} 
found that this correction
can be schematically written as 
\begin{equation}
\label{eq:glmpred}
\mathcal{P}_{\zeta}/\mathcal{P}_0\simeq  1+\mathcal{F}_l+\mathcal{F}_h+\mathcal{F}_{lh} \, ,
\end{equation}
and, for a turn of negligible duration
and a mass parameter of the heavy field much larger than $H$,
they found $\mathcal{F}_l=0$
and $\mathcal{F}_h\sim\Delta\theta^2$,
while the contribution $\mathcal{F}_{lh}$ was left 
unevaluated\footnote{The definition of the correction factors
$\mathcal{F}_l$, $\mathcal{F}_h$ and $\mathcal{F}_{lh}$
is given in the Appendix.}.
Given the disparity of
these results and the fact that they 
predict or suggest a non-vanishing effect for all modes with $k>k_\mathrm{turn}$,
it is interesting to take a closer look at the evolution of the perturbations at the turn. 
As we will argue at the end of this section, 
the principal object of interest
 is the full expression for the power spectrum, and therefore
 each of the individual contributions to its formula 
 should be carefully tracked, since it is possible that 
 cancellations might occur. In particular 
 such care will be essential
 for the scales we want to focus on in this paper
 when the turn occurs.

Our approach to this
problem will be two-fold: we first discuss analytically 
the solutions to the equations of motion for the perturbations 
in the sub-horizon regime,
and then 
corroborate our findings  
in comparison to numerical 
estimates  
for a simple model.
Our estimates will not rely 
on the effective field theory treatment
because the sharpness of the turn can be sufficient to induce 
oscillations of the heavy field, which  
can make the 
effective field theory description break down. Instead, 
we proceed with the analysis of the perturbations for both
fields directly, 
keeping both degrees of freedom (light and heavy fields)
manifest throughout the calculation.

\subsection{Evolution of the perturbations at the turn: analytical discussion}
\label{sec:f1a}
In multiple field modes the dynamical evolution
is sometimes best described in terms of curvature (adiabatic) 
and isocurvature (entropy) perturbations \cite{Gordon:2000hv}.
These can be obtained by performing an instantaneous 
rotation with respect to the field perturbations themselves, 
and can be interpreted as the perturbations defined along and 
perpendicularly to the inflationary trajectory.
We can then introduce the 
gauge-invariant 
Mukhanov--Sasaki variables \cite{Mukhanov:1985rz,Sasaki:1986hm}, 
which we shall denote by
$Q_\sigma$ and $\delta s$, 
respectively.
On super-Hubble scales, these 
represent the instantaneous curvature and isocurvature perturbations,
$\zeta=Q_\sigma/\sqrt{2\varepsilon}M_P$ and $\mathcal{S}=\delta s/\sqrt{2\varepsilon}M_P$.

The general equations of motion for these variables are usually
rather lengthy, 
and their derivation can be found, for example, in Ref.~\cite{Lalak:2007vi}.
They can, however, be significantly simplified under
the assumption that the slope and the curvature of the inflationary potential
change negligibly 
around  
the  
portion of the inflationary trajectory
where the turn takes place.
We also assume that at a given instant the trajectory momentarily changes 
direction, 
and follows a straight line in field space after the turn.
Moreover, we only consider models for which $\varepsilon$
is very small and nearly constant.

Subject to these simplifying approximations, 
the equations of motion for 
the
rescaled comoving perturbations, 
$v_\sigma = a\,Q_\sigma$ and $v_s=a\,\delta s$,
assume a much simpler form.
Up to slow-roll corrections, they can be written as
\begin{eqnarray}
\label{eq:eom1}
v_\sigma'' + \left(k^2+\frac{\mu_{\sigma}^2-\rho^2-2}{\tau^2}\right) v_\sigma + 
 \left(\frac{2\rho}{\tau}v_s\right)' - \frac{2\rho}{\tau^2}v_s = 0 \\
\label{eq:eom2}
v_s'' + \left(k^2+\frac{\mu_{s}^2-\rho^2-2}{\tau^2}\right) v_s 
-\frac{2\rho}{\tau} v_\sigma'-\frac{2\rho}{\tau^2}
v_\sigma = 0 \ .
\end{eqnarray}
Above we have parametrised the rate of turn by introducing
\begin{equation}
\label{eq:rhodef}
\rho\equiv\frac{\dot\theta}{H}= \frac{\mathrm{d}\theta}{\mathrm{d}N} \ ,
\end{equation}
where $N=\int{H\d t}$ is the number of e-folds. 
Away from the turn $\rho$ is approximately zero.
In Eqs. \eqref{eq:eom1} and \eqref{eq:eom2}
$\mu_{\sigma,s}^2$ are the mass parameters
of the perturbations, given in Hubble units, which can be read from the full equations of motion.
We shall assume that the mass of the variable associated 
with the curvature perturbation $v_\sigma$, 
denoted by $\mu_\sigma$, 
is negligible in comparison
to the Hubble scale,$\mu_\sigma^2\ll1$. We notice the presence of a gravitational mass (squared) term, $-2/\tau^2$, 
which results from the dynamical spacetime background. Such term would be
absent in Minkowski spacetime.

In the following, we parametrize the localised sharpness of the turn
by assigning $\rho$ a Dirac delta distribution, as done by Shiu \& Xu \cite{Shiu:2011qw}.
 This is an appropriate description of 
 the trajectory's feature: it starts with a straight line
 in field space and rapidly changes its direction to
 continue as a straight line after the turn. More specifically, we assume
that the turn rate, $\theta'$, can be 
                            described by
\begin{equation}
\label{eq:delta}
\theta'(\tau)=\Delta\theta\,\delta(\tau-\tau_\mathrm{turn}) \,,
\end{equation}
where  
$\tau_\mathrm{turn}$ 
corresponds to the value of the conformal time 
at the turn. 
In our analysis we will be mainly interested in the dynamics of modes
which are still sub-horizon 
when the turn happens.
Therefore, we can restrict our attention to the case $|k\tau_\mathrm{turn}|\gg\mu_s$,
for which the mass parameters $\mu_\sigma$ and $\mu_s$, 
as well as the gravitational mass term, 
are 
comparatively negligible.

It appears that the ansatz (\ref{eq:delta}) is problematic
as in Eqs.~(\ref{eq:eom1}) and (\ref{eq:eom2}) we can find a square
and a derivative of $\theta'$. This is, however, just an artifact of 
working with a time-dependent basis 
for the perturbations.\footnote{This problem does not appear if the delta-like singularity
can be somehow resolved. Ref.\ \cite{Noumi:2013cfa} gives an excellent example of such a solution; however, it follows 
from Eqs.~(\ref{eq:eom1}) and (\ref{eq:eom2})
 that working in the curvature-isocurvature (`kinetic') basis does not remove the singularity.}
Indeed, $v_\sigma$ and $v_s$ represent perturbations along and perpendicular to the inflationary trajectory, respectively, 
a notion that becomes somewhat ill-defined at the instantaneous turn.

Therefore, it will be helpful to 
rewrite
Eqs.\ (\ref{eq:eom1}) and (\ref{eq:eom2}) 
by performing a field rotation such that the
new perturbations correspond to fixed directions in  
field space.
A general rotation by a constant angle $\theta_0$ reads
$\vec v=\mathbb{R}\vec u$, where $\vec v=(v_1,v_2)^T$, $\vec u=(u_1,u_2)^T$ and 
$\mathbb{R}$ is a time-dependent rotation matrix 
\begin{equation}
\mathbb{R} = \left( \begin{array}{cc} \cos(\theta-\theta_0) & -\sin(\theta-\theta_0) \\ \sin(\theta-\theta_0) &  \cos(\theta-\theta_0) \end{array}\right) \ .
\end{equation}
Since this matrix satisfies
\cite{Lalak:2007vi,GrootNibbelink:2001qt}
\begin{equation}
\label{eq:Rprime}
\mathbb{R}' = \frac{\rho}{\tau} \ \mathbb{E} \ \mathbb{R}\, ,
\end{equation}
where $\mathbb{E}$ is an antisymmetric $2\times2$ matrix 
with $\mathbb{E}_{21}=1$, 
then Eqs.\ (\ref{eq:eom1}) and (\ref{eq:eom2}) 
expressed in terms of $\vec u$ become
\begin{equation}
\label{eq:eom3}
\vec u'' +\left(k^2 - \frac{1}{\tau^2}\mathbb{R}^T \, \mathbb{Q} \,\mathbb{R} \right) 
\vec u=0\,,
\end{equation}
in which 
\begin{equation}
\mathbb{Q} = \left( \begin{array}{cc} 0 & 3\rho - \rho'\tau \\ 3\rho - \rho'\tau & 0 \end{array}\right)\ .
\end{equation}
We note that the passage to the new variables $\vec u$ made
potentially troublesome terms proportional to $\rho^2$ in 
Eqs. (\ref{eq:eom1}) and (\ref{eq:eom2}) vanish, and that
the effective `mass matrix' of the perturbations, $\mathbb{R}^TQ\mathbb{R}$, 
is now symmetric.
Using $\frac{3\rho}{\tau^2}-\frac{\rho'}{\tau} = \tau^2\frac{\mathrm{d}}{\mathrm{d}\tau} \frac{\theta'}{\tau^2}$, we can cast Eq. (\ref{eq:eom3}) in the form
\begin{equation}
\label{eq:eom4}
\vec u'' + \left[ k^2-\tau^2\frac{\mathrm{d}}{\mathrm{d}\tau}
\left(  \frac{\theta'}{\tau^2} \right) \mathbb{R}_2 \right] \vec u = 0\,,
\end{equation}
with the orthogonal matrix $\mathbb{R}_2$ being given by
\begin{equation}
\mathbb{R}_2 = \mathbb{R}^T\, 
\left( \begin{array}{cc} 0 & 1 \\ 1 & 0 \end{array}\right) \,
       \mathbb{R}
      =
      \left( \begin{array}{cc} \sin 2(\theta-\theta_0) & \cos 2(\theta-\theta_0) \\
      \cos 2(\theta-\theta_0) & -\sin 2(\theta-\theta_0) \end{array}\right) \, .
\end{equation}

In the equation above there remains the 
subtlety of  
having to deal with the derivatives of the Dirac delta distribution. 
This issue is 
solved when we rewrite the components of Eq. (\ref{eq:eom4}) as
\begin{eqnarray}
\label{eq:eom4a}
u''_1  - \theta'' \left[\sin2(\theta-\theta_0)\, u_1+\cos2(\theta-\theta_0)\,u_2\right]+\ldots &=& 0 \\
\label{eq:eom4b}
u''_2  - \theta'' \left[\cos2(\theta-\theta_0)\, u_1-\sin2(\theta-\theta_0)\,u_2\right]+\ldots &=& 0 \, ,
\end{eqnarray}
where the ellipses represent terms which do not affect regularity of the solutions.
Integrating Eqs.~(\ref{eq:eom4a}) and (\ref{eq:eom4b}) over an infinitesimally small interval
$(\tau_-,\tau_+)$
around $\tau_\mathrm{turn}$,
and applying the identity
$\int_{\tau_-}^{\tau_+}\delta'(\tau)f(\tau)\equiv 
-\int_{\tau_-}^{\tau_+}\delta(\tau)f'(\tau)$,
we find that there are solutions corresponding to a vanishing jump of 
$\vec u'_1$ and $\vec u'_2$ across $\tau_\mathrm{turn}$.
If the rate of turn is the largest scale in the
dynamics, 
these are
\begin{equation}
\label{eq:mc2a}
\vec u^{(1)} = \left( \begin{array}{c} \cos 2(\theta-\theta_1) \\ -\sin 2(\theta-\theta_1) \end{array} \right) 
\qquad\textrm{and}\qquad
\vec u^{(2)} = \left( \begin{array}{c} \sin 2(\theta-\theta_1) \\ \cos 2(\theta-\theta_1) \end{array} \right) \, ,
\end{equation}
where $\theta_1$ is an arbitrary constant. A choice of $\theta_1=\theta_0$
is particularly convenient, as it corresponds to $u^{(J)}_I=\delta^J_I$ 
as the initial condition.
This gives
\begin{equation}
\label{eq:mc2b}
\vec v^{(1)} = \left( \begin{array}{c} \cos \theta \\ -\sin \theta \end{array} \right) 
\qquad\textrm{and}\qquad
\vec v^{(2)} = \left( \begin{array}{c} \sin \theta \\ \cos \theta \end{array} \right) \, .
\end{equation}
Hence, the main effect of a sudden turn of the inflationary trajectory consists in a rotation of the components of the perturbations for a range of sub-horizon modes. Since such an
orthogonal transformation leaves the correlation functions invariant, we conclude that 
after the turn the final power spectrum of the curvature perturbations $\mathcal{P}_{\zeta}$
is {\it equal} to $\mathcal{P}_{\zeta 0}$, the value that the power spectrum would have attained in the absence of a turn.

We have so far dealt with the case in which $\theta'$ is the dominant scale in the problem. It is much simpler to
solve Eqs.~(\ref{eq:eom1}) and (\ref{eq:eom2}) for very short wavelengths, still assuming a fast turn, 
$k/a_\mathrm{turn}\gg \theta' \gg \left\{\mu_s, \mu_\sigma\right\}$. 
In this case, we can neglect all terms which do not contain a derivative of a
wavefunction in 
Eqs.~(\ref{eq:eom1}) and (\ref{eq:eom2}), except for the
gradient terms 
\begin{eqnarray}
\label{eq:eom1v2}
v_\sigma'' + k^2 v_\sigma -2\theta'v'_s  = 0 \\
\label{eq:eom2v2}
v_s'' +k^2 v_s +2\theta' v_\sigma' = 0 \, .
\end{eqnarray}
It is easy to check that the approximate solutions to Eqs.~(\ref{eq:eom1v2}) and (\ref{eq:eom2v2}) read
\begin{equation}
\label{eq:mc2c}
\vec v^{(1)} = \left( \begin{array}{c} \cos \theta \\ -\sin \theta \end{array} \right) e^{-\imath k\tau}
\qquad\textrm{and}\qquad
\vec v^{(2)} = \left( \begin{array}{c} \sin \theta \\ \cos \theta \end{array} \right) e^{-\imath k\tau} \, .
\end{equation}
Again we can see that the main effect of a fast turn consists in an orthogonal transformation of the modes.
Hence the power spectrum {\it does not change} with respect to the single-field slow-roll case. This conclusion
is not surprising---during the turn the large-$k$ modes are deep inside the Hubble and mass radii, and they behave
as massless fields in Minkowski space.
Note that this is different from what happens
with respect to modes which are super-horizon 
when the turn takes place.\footnote{ 
A notable example is that of double quadratic inflation 
(see, for example, Vernizzi \& Wands \cite{Vernizzi:2006ve}).
It is well known that, in this case, 
there are appreciable changes to the power spectrum, 
as well as other observables, such as the scalar spectral index
and local non-gaussianity as measured by $\fNL$ 
(see, for example, Dias \& Seery \cite{Dias:2011xy}). 
}

\subsection{Evolution of the perturbations at the turn: numerical study}
\label{sec:f1b}

\begin{figure}
\begin{center}
\includegraphics*[width=10cm]{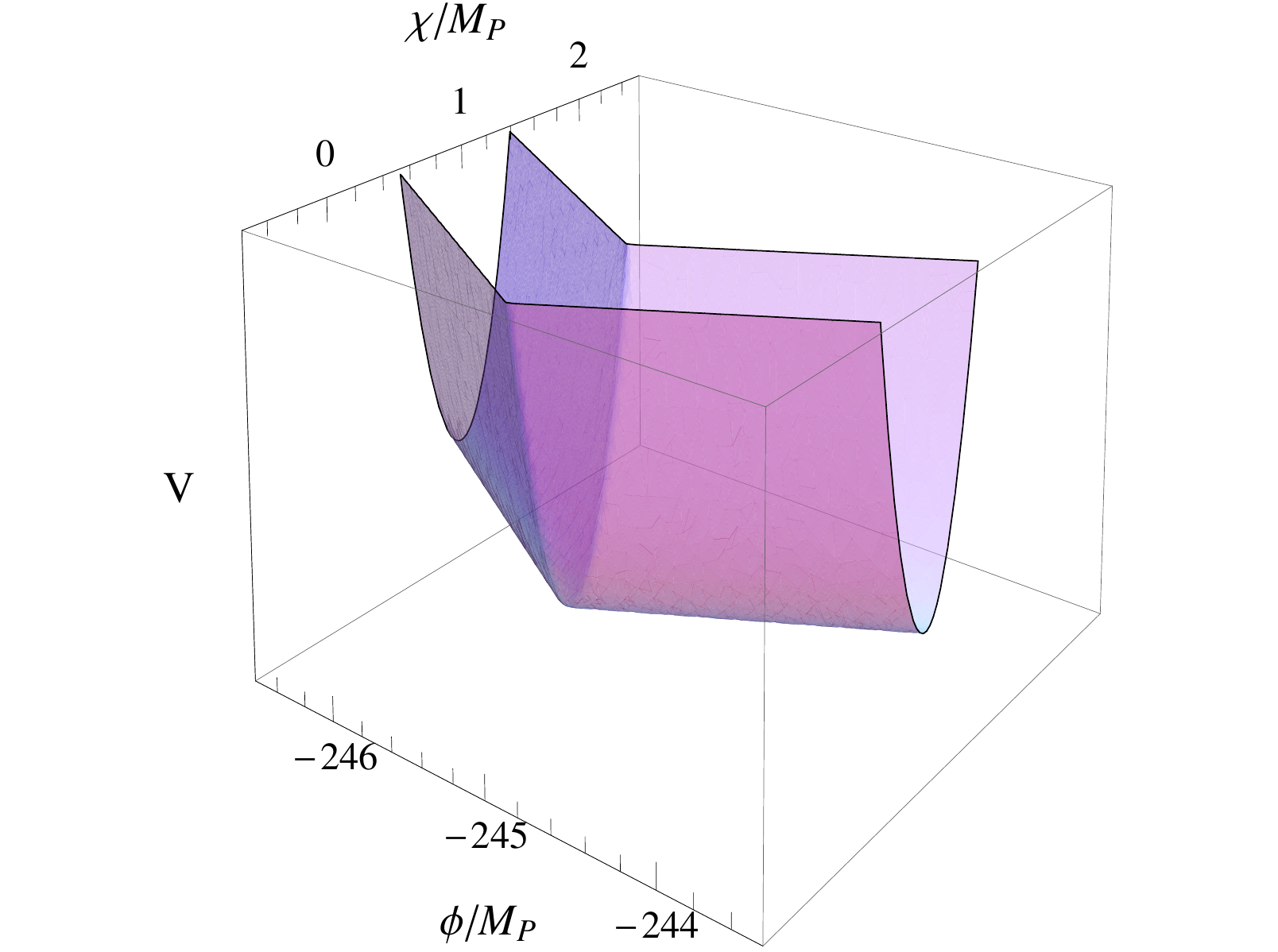}
\end{center}
\caption{\sl A sketch of the potential (\ref{eq:pot}) with the parameter choice (\ref{eq:potpar}) (except 
that the value $\Delta\theta=\frac{\pi}{4}$ was chosen
to visually make the turn more pronounced).
 \label{f_pot}}
\end{figure}

Gao, Langlois \& Mizuno \cite{Gao:2012uq} 
put forth a simple two-field potential which allows
studying the effects of a sudden turn of the inflationary trajectory. 
The potential between the two 
canonically normalised fields, $\phi$ and $\chi$, reads
\begin{equation}
\label{eq:pot}
V(\phi,\chi) = \frac{1}{2}m_\phi^2\,\phi^2 + \frac{1}{2}M^2\cos^2\left(\frac{\Delta\theta}{2}\right) \Big[ \chi - (\phi-\phi_0)\tan\Xi(\phi)\Big]^2 \, ,
\end{equation}
with $\Xi(\phi)=\Delta\theta\,\mathrm{arctan}\left[s(\phi-\phi_0)\right]$. Here, $\Delta\theta$ is a constant parameter corresponding
to the net angle variation during the turn, 
while adjusting the 
constant parameter $s$  
allows to control the sharpness of the turn. In the following,
we fix the parameters of the potential as
\begin{equation}
\label{eq:potpar}
m_\phi = 10^{-7}\, M_P\ , \
M=2\times 10^{-4}M_P\ , \
 \phi_0=-100\sqrt{6} \, M_P\ , \
 \delta\theta=\frac{\pi}{10}\ ,  \
 s=5000\sqrt{3} \, M_P^{-1} \, ,
\end{equation}
in which we reintroduced the Planck mass for clarity.
The shape of the potential corresponding to this choice is shown in Figure \ref{f_pot}. Indeed, it corresponds to a slowly descending
valley with steep slopes exhibiting a  
turn. With the parameters choice (\ref{eq:potpar}), 
the mass of the heavy perturbation
(or equivalently, the curvature of the potential across the bottom of the valley)
 is twenty times larger than the Hubble parameter. It is in this sense
 that we can refer to the turn as \emph{sharp}. 

In order to discuss the evolution of the perturbations numerically, it is convenient to adopt yet another time-dependent 
basis for the perturbations. Following 
the same prescription as Gao, Langlois \& Mizuno, 
we  
will track the evolution of the perturbations, 
schematically denoted by $\varphi_I$,
in the directions determined by the eigenvectors of the mass matrix 
\cite{Mukhanov:1985rz,Sasaki:1986hm,Sasaki:1995aw}
\begin{equation}
\mathbb{M}_{IJ} = 
V_{,IJ}-\dfrac{1}{a^3} \dfrac{\d}{\d t}
\left(\dfrac{a^3}{H} \dot\varphi_I \dot\varphi_J\right)=
V_{,IJ} + (3-\varepsilon)\, \dot\varphi_I\dot\varphi_J+\frac{1}{H} \left( V_{,I}\, \dot\varphi_J+V_{,J}\, \dot\varphi_I\right) \, ,
\end{equation}
where $V_{,I}= \frac{\partial V}{\partial\varphi_I}$
and $V_{,IJ}= \frac{\partial^2 V}{\partial\varphi_I\partial\varphi_J}$.
The mass matrix is related with the expansion 
tensor of the inflationary flow via 
$\mathbb{U}_{IJ}=-\mathbb{M}_{IJ}/3H^2$ \cite{Seery:2012vj}, 
and it is often more
useful to work with its eigenvalues.
Indeed, denoting by $v_l$ and $v_h$ the perturbations corresponding to the smaller and larger mass eigenvalues, $m_l^2$ and $m_h^2$, respectively,
we obtain the following equations of motion
\begin{eqnarray}
\label{eq:glmeom1a}
v''_l + \left( k^2+a^2m_l^2-\vartheta'^2-\frac{a''}{a} \right) v_l -\vartheta''v_h-2\vartheta'v'_h &=& 0 \\
\label{eq:glmeom1b}
v''_h + \left( k^2+a^2m_h^2-\vartheta'^2-\frac{a''}{a} \right) v_h +\vartheta''v_l+2\vartheta'v'_l &=& 0 \, ,
\end{eqnarray}
where $\vartheta$ is the angle of the rotation 
that diagonalizes $\mathbb{M}$.
Since by assumption $|\vartheta'|\to0$ 
in the regime of very early and very late times 
(well before and well after the turn), at late times the light and the heavy modes
naturally correspond to the curvature and the isocurvature perturbations, respectively.

\begin{figure}
\begin{center}
\includegraphics*[width=10cm]{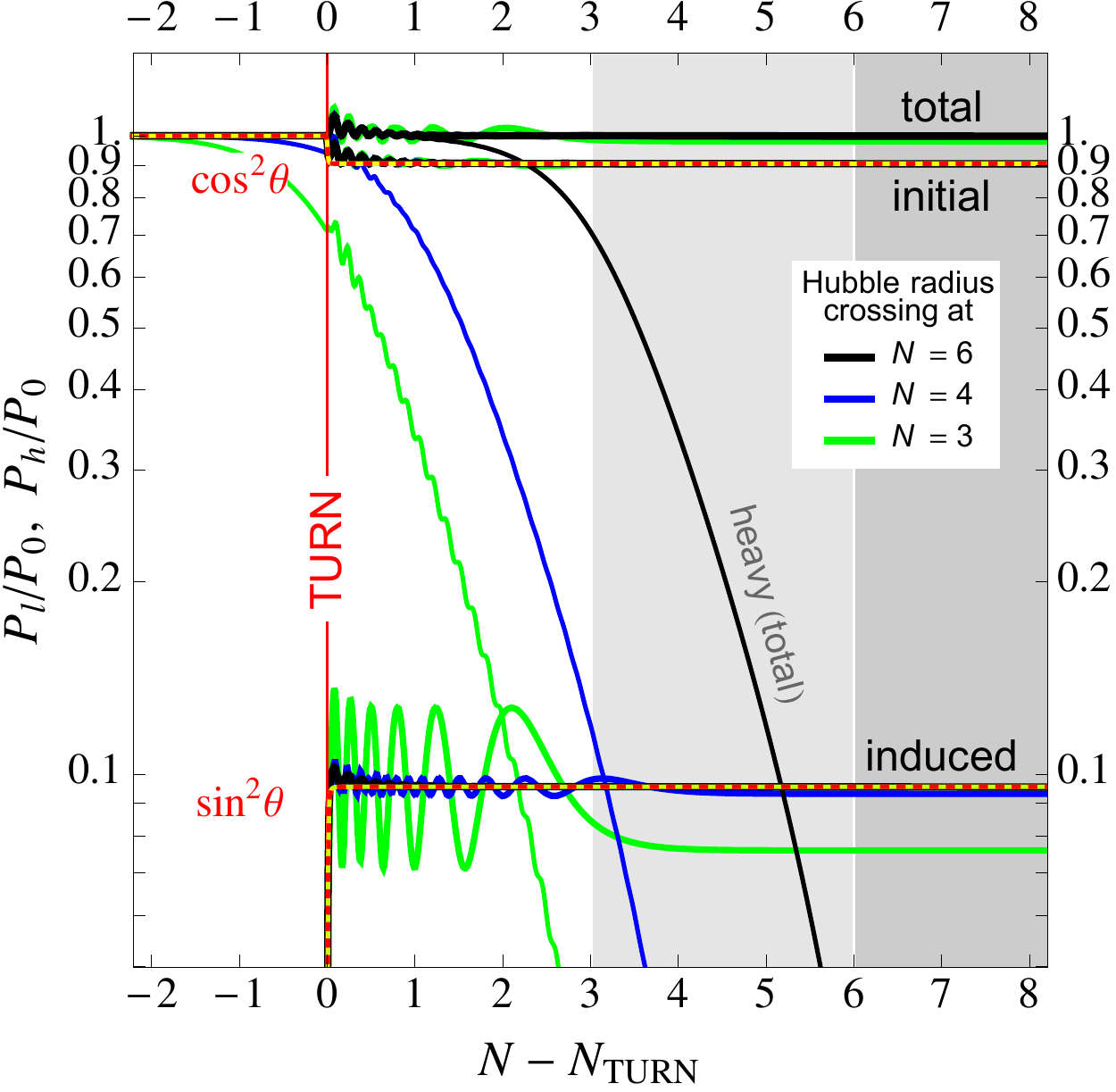}
\end{center}
\caption{\sl The evolution of the power spectra 
$\mathcal{P}_l$ and $\mathcal{P}_h$ 
of the light and the heavy modes, respectively, that leave the Hubble radius 
3, 4 and 6 e-folds after
the turn. 
Lighter (darker) shaded regions denote times at which the 
largest-$k$ of these modes is outside the mass radius $k=am$ (Hubble radius $k=aH$).
Different lines of the same colour correspond to
the various contributions to the 
light mode (the perturbation that remained of the light-only initial condition, 
the perturbation that
was induced by the heavy-only initial condition and the total perturbation, 
defined as the sum of squares of
amplitudes of the components), 
or to the total power spectrum of the heavy mode.
Dashed lines show the behaviour of $\cos^2\vartheta$ and
$\sin^2\vartheta$. 
 \label{f_pert1}}
\end{figure}

We solve Eqs.~(\ref{eq:glmeom1a}) and (\ref{eq:glmeom1b})
assuming the initial 
state of the curvature and isocurvature perturbations
is the Bunch--Davies vacuum. For clarity, we focus on
a
single mode of wavenumber $k$ 
much larger than the wavenumber $k_\mathrm{turn}$ of the mode
that leaves the Hubble radius at the 
turn---namely, we take the modes that leave the Hubble radius
3, 4 and 6 e-folds after the turn.
The results are shown in Figure \ref{f_pert1}.  
As our initial conditions are $(v^{(1)}_l,v^{(1)}_h)\propto(1,0)$ and
$(v^{(2)}_l,v^{(2)}_h)\propto(0,1)$, 
we refer to $v^{(1)}_l$ as the {\it initially} light perturbation,
while 
$v^{(2)}_l$ is the {\it induced} light perturbation; 
the {\it total} light perturbation arises from
adding these contributions in quadratures. 
Similar nomenclature is applied to refer to the heavy
field 
perturbations. 

In Figure \ref{f_pert1} we show for several modes the 
initial, 
induced 
and the total light perturbation, as well as the total heavy perturbation.
The  
instantaneous power spectra are normalised to $\mathcal{P}_0(\tau)$,
which is the instantaneous power spectrum of a single
massless mode with
elementary wavefunction 
$v_k(\tau)=\frac{e^{-\imath k\tau}}{\sqrt{2k}}\left(1-\frac{\imath}{k\tau} \right)$, 
corresponding to
the same values of the Hubble parameter and the field velocity at the Hubble radius exit.
Figure \ref{f_pert1}  
shows that the evolution of the modes closely follows the estimate
(\ref{eq:mc2b})---this estimate works well 
if the turn happens when the mode is still within its mass horizon, 
for which
$k=am_h$. Small wiggles appearing immediately after the turn result from the fact that
it takes a small finite amount of time before the true 
inflationary trajectory settles at the bottom of the
potential valley.
We also checked that the deviations of $\mathcal{P}_l$ from $\mathcal{P}_0$
do not exceed 1\% for the modes which are within their mass horizons at the turn; this conclusion
does not change if the mass of the heavy field is increased or reduced by a factor of 2.

\para{Comparison with previous results}
At this point, it is relevant
to understand why our result is 
different from
the corrections reported in Refs.~\cite{Shiu:2011qw,Gao:2012uq}.
The authors of Ref.~\cite{Shiu:2011qw} used equations of motion
different from Eqs. (\ref{eq:eom1}) and (\ref{eq:eom2}),  
which results in  
a different estimate of the correction to the power
spectrum compared to numerical estimates.

On the other hand, the authors of Ref.~\cite{Gao:2012uq} 
based their estimate on
a perturbative calculation, with $\vartheta'$ as the expansion
parameter. 
Applying our approximation of 
                $\vartheta'$ as a Dirac delta distribution in the
                formalism of Ref.~\cite{Gao:2012uq} 
                would lead to singularities due to the presence of
                terms involving
                $\vartheta'^2$ in integrals.
To circumvent this problem, we regularize our treatment of the fast turn,
choosing the profile of $\vartheta'$ either as a Gaussian or a product
of two step functions. The respective formulae and the resulting expressions
for $\mathcal{F}_l$, $\mathcal{F}_h$ and $\mathcal{F}_{lh}$ are given in Table \ref{tabf}.
To obtain these results,  we have used the fact that the
wave functions
for the modes  
inside the mass or Hubble horizon can
be expressed as 
$v_{l,h}\approx\frac{1}{\sqrt{2k}}e^{-\mathrm{i}k\tau}$.
This is justified since we are assuming a 
perturbative correction to the power 
spectrum and therefore it suffices to use the lowest-order 
result for the wavefunction.
It is clear that for both parametrizations of $\vartheta$ 
the corrections
$\mathcal{F}_l$ and $\mathcal{F}_{lh}$ diverge in the fast-turn limit when $\Delta\tau\to0$, but
the sum of the three corrections entering Eq.~(\ref{eq:glmpred}) is zero,which is in effect 
    the physical estimate for the corrections to the power spectrum.
Hence the actual prediction of Ref.~\cite{Gao:2012uq} for the small-scale 
power spectrum is that the power spectrum is to a good accuracy indistinguishable
from one obtained in the single-field case.\footnote{One can try to estimate
how fast the effects of the turn decouple, by expanding the wave function of the
heavy mode in powers of the finite mass corrections. It is easy to check that
the leading correction is $\mathcal{O}(\frac{M^2}{H^2}\frac{k_\mathrm{turn}^2}{k^2})$,
i.e.~it decreases as $k^{-2}$. 
}
This conclusion complements calculations performed by Noumi \& Yamaguchi \cite{Noumi:2013cfa},
who studied analytically and numerically the evolution of the perturbations with $H<k/a<M$; their
numerical results also suggest the null effects of the turn for large $k$.

\begin{table}

	\tablepreamble
		
	\sbox{\tableA}{%
		\begin{tabular}{q|Qq}
			 	\toprule			
			\multicolumn{1}{c|}{function}
			& \multicolumn{2}{c}{\textbf{$\vartheta'$ profiles}}
		 	\\ 
\multicolumn{1}{c|}{}
		 	&
		 	\multicolumn{1}{c|}{$\frac{1}{\sqrt{\pi}}\frac{\Delta\vartheta}{\Delta \tau} \,\mathrm{exp}\!\left[-\frac{(\tau-\tau_\mathrm{turn})^2}{\Delta\tau^2}\right]$} &
		 	\multicolumn{1}{c}{$\frac{\Delta\vartheta}{\Delta \tau} \Theta(\tau-\tau_\mathrm{turn})\Theta(\tau_\mathrm{turn}+\Delta\tau-\tau)$} 		 \vspace*{0.3cm}	
		 	\\

			\midrule

			\mathcal{F}_l & \frac{\Delta\vartheta^2}{\sqrt{2\pi}k\,\Delta\tau} e^{-\frac{1}{2}k^2\Delta\tau^2}\sin\left(2k/k_\mathrm{turn}\right)&
			 \Delta\vartheta^2 \left[ \frac{\sin\left(2k/k_\mathrm{turn}\right)}{k\,\Delta\tau}+\cos\left(2k/k_\mathrm{turn}\right)\right] 			\\

			\cmidrule{1-3}

			\mathcal{F}_h & \Delta\vartheta^2 & \Delta\vartheta^2 			\\

			\cmidrule{1-3}

			\mathcal{F}_{lh} & -\Delta\vartheta^2 -\frac{\Delta\vartheta^2}{\sqrt{2\pi}k\,\Delta\tau} e^{-\frac{1}{2}k^2\Delta\tau^2}\sin\left(2k/k_\mathrm{turn}\right) & -\Delta\vartheta^2\left[ 1 +  \frac{\sin\left(2k/k_\mathrm{turn}\right)}{k\,\Delta\tau}+\cos\left(2k/k_\mathrm{turn}\right)\right] \\

			\bottomrule

		\end{tabular}
	}
	\settowidth{\tblw}{\usebox{\tableA}}
	\addtolength{\tblw}{-1em}

	\begin{center}
		\usebox{\tableA}
	\end{center}

	\renewcommand{\arraystretch}{1}

	\caption{\sl Corrections $\mathcal{F}_l$, $\mathcal{F}_h$, $\mathcal{F}_{lh}$ to the power spectrum
calculated for two different parametrizations of the turn rate $\vartheta'$.   \label{tabf}}
	\end{table}


\section{Particle production at sharp turns} \label{sec:pp}
So far we have analysed the dynamics of the field perturbations 
when they encounter a turn in the classical trajectory in field configuration
space. For the preceding analysis we have dismissed the possibility of 
producing particles and the 
induced displacement of the heavy field from its local vacuum, 
as a consequence of the turn.
However, if the turn is \emph{non-adiabatically} sharp
 we should expect particle production to
be effective.
In this section, we shall focus on such an event of particle production.

\subsection{Particle production in  
single-field inflation}
\label{sec:pp1}

Before proceeding with the analysis of production of quanta of particles
induced by the turn, 
it is instructive to recapitulate
the issue of particle production in 
a simpler case.  We follow the 
work by Miji\'{c} \cite{Mijic:1998if}, 
and 
consider a model with 
a single, massive, canonically normalised scalar field, $\varphi$,
in a de Sitter background ($H=\mathrm{const}$). 
Notice that giving the scalar field a mass, 
breaks conformal 
invariance and introduces  
a scale in the problem
to be compared to the scale associated with horizon exit. 

The equation of motion for the elementary comoving
wavefunction $u_k(\tau)=a(\tau)\varphi_k(\tau)$ 
associated with this field has the well-known 
form\footnote{From now on we shall drop the  
subscript $k$, 
unless it is necessary to distinguish between different modes.}
\be
\label{eq:eomp1}
0 =u''+\left(k^2+m^2a^2-\frac{a''}{a}\right)u=
u''+\left[ k^2 + \frac{1}{\tau^2}\left(\frac{m^2}{H^2}-2 \right)\right]u\, ,
\ee
where we used the de Sitter relations $aH\tau=-1$
with constant $H$,
which is a good approximation if $\varepsilon\ll1$.
The solution to Eq. (\ref{eq:eomp1}) reduces in 
the early times regime, $\tau\to-\infty$, 
to a positive-frequency plane-wave
\be
\label{eq:uhankel}
u(\tau) = e^{\imath\pi(\nu-1/2)/2}\ \frac{\sqrt{\pi}}{2} \, 
\sqrt{-k\tau} \, H^{(1)}_\nu(-k\tau) \, ,
\ee
where $H^{(1)}_\nu$ is the Hankel function of the
first kind and order $\nu=\sqrt{\frac{9}{4}-\frac{m^2}{H^2}}$.

This function can be decomposed 
using a Bogolyubov transformation, also known as 
the \emph{mode mixing prescription}, as follows
\be
\label{eq:bogo1}
u = \frac{1}{\sqrt{2\omega}}(\alpha+\beta) \, .
\ee
The Bogolyubov coefficients 
$\alpha$ and $\beta$ stand for the WKB-like positive and negative frequency modes
\be
\label{eq:bogo1a}
\alpha = A\, \mathrm{exp}\left[ -\imath\int_{}^{\tau} \omega(\eta)\,\mathrm{d}\eta\right]\qquad\textrm{and}\qquad
\beta = B\, \mathrm{exp}\left[ +\imath\int_{}^{\tau} \omega(\eta)\,\mathrm{d}\eta\right]\ \ ,
\ee
and
the dispersion relation reads
\be
\omega^2 = k^2 + \frac{1}{\tau^2}\left(\frac{m^2}{H^2}-2 \right) \, .
\ee
The functions $\alpha$ and $\beta$ obey a system of coupled differential equations, 
which determines their time evolution, as follows:
\begin{equation}
	  \left\{
		\begin{array}{l@{\hspace{5mm}}l}
\alpha' =& -\imath \omega \alpha + \dfrac{\omega'}{2\omega}\beta \\	\\		
\beta' =& \ \ \, \imath \omega \beta + \dfrac{\omega'}{2\omega}\alpha \ .
		\end{array}
		\right. 
		\label{eq:eomab}
	\end{equation}
Provided $\omega^2>0$, they satisfy the Wronskian 
condition $|\alpha|^2-|\beta|^2=1=|A|^2-|B|^2$.

It may appear that Eqs.~(\ref{eq:eomab})  
become singular
for $\omega\to0$. However, there is a 
simple prescription that allows a determination of $\alpha$ 
and $\beta$ once the solution for the wavefunction $u$ is known.
Indeed,  
using Eqs. (\ref{eq:bogo1}) and \eqref{eq:eomab}, 
we find
\begin{equation*}
u' = -\imath\sqrt{\frac{\omega}{2}} \ (\alpha-\beta)\, .
\end{equation*} 
Hence
\begin{equation}
\alpha = \frac{1}{\sqrt{2}}\left( \sqrt{\omega} \, u+
\frac{\imath}{\sqrt{\omega}} \, u'\right) \ \ \ \textrm{and} \ \ \ 
\beta= \frac{1}{\sqrt{2}}\left( \sqrt{\omega}\, u
-\frac{\imath}{\sqrt{\omega}} \, u'\right) \, .
\label{eq:bet1}
\end{equation}

The factor $|\beta|^2$, which can 
usually be interpreted as the occupation number of a given 
mode\footnote{Certainly, utmost care is required 
                when one interprets $|\beta|^2\neq 0$ as the presence of 
                real particles \cite{Birrell:1982ix,Linde:2005ht} 
                in time-dependent backgrounds. 
                We will leave the discussion of 
                the ambiguities associated with this interpretation
                for \S \ref{sec:particles}, which are nevertheless 
                of no consequence to our results. 
                In particular, we consider a very short non-adiabaticity period, 
                and focus on modes deep inside the Hubble radius 
                (for which the Minkowski spacetime approximation works very well).}
reads, for $\omega^2>0$ \cite{Garbrecht:2002pd}
\be
\label{eq:ngps}
|\beta|^2 = \frac{|u'|^2+\omega^2|u|^2}{2\omega} - \frac{1}{2} \, .
\ee
Far outside the horizon, we can expand 
Eq. (\ref{eq:uhankel}) for $k|\tau|\ll 1$, 
            to find\footnote{ Here we used the asymptotic 
                              expansion of the 
                              Hankel function in \eqref{eq:uhankel}, 
                              which for a massive field
                              corresponds to an
                              imaginary order $\nu=\imath m/H$, and
                              gives in the limit when $x\ll 1$
                              (that is, very late times)
                              \begin{equation*}
                              H_{\nu}^{(1)} (x)\rightarrow
                              -(1+\imath) \,
                              \frac{2^{\imath m}}{\sqrt{m\pi}}\,
                              e^{-m(\imath +\pi/2-
                              \imath \ln m)} \, 
                              (-k \tau)^{-\imath m}\ . 
                              \end{equation*}
                              }           
\be
u_k (\tau,m) \approx \left\{ \begin{array}{ll} 
-\dfrac{\imath}{\sqrt{2}k^{3/2}\tau} \ , & \textrm{for }\  \dfrac{m}{H}\ll 1
\\ \\
\sqrt{-\dfrac{H\tau}{2m}} \mathrm{exp}
\left[\dfrac{\imath m}{H}\Big(\ln\left(-\frac{Hk\tau}{2m}\right)+1\Big)\right]
\ , & \textrm{for } \ \dfrac{m}{H}\gg 1\ .
   \end{array} \right.
   \label{eq:umassive}
\ee
Substituting into
 Eq. (\ref{eq:ngps}),
 we  
 observe that 
 if the field is very light compared to the Hubble scale, $m/H\ll 1$, the occupation
number  
grows like $\sim (-k\tau)^{-3}$ after  
Hubble radius 
crossing\footnote{Ref.~\cite{Garbrecht:2002pd} uses a Wigner function approach and 
finds a different, yet still divergent, dependence;
the Lagrangian used for quantization there contains mixed 
$uu'$ terms, in contrast with our presentation.}---the field fluctuations
become semiclassical. On the other hand, for large $m/H$ 
we obtain the occupation number approaching a constant value
\be
\lim_{\tau\to 0^-}|\beta|^2 = \frac{H^2}{16m^2}\ll1 \, ,
\ee
indicating that a heavy field does not form a semiclassical condensate, 
in agreement with the findings of Haro \& Elizalde \cite{Haro:2008zz}.

\subsection{Generalisation to the multiple-field case}
\label{ss1}

The formalism outlined above can be readily applied to the case of multiple-field inflation.
Following  Nilles, Peloso \& Sorbo \cite{Nilles:2001fg}, 
we now consider a system of $n$ canonically normalised scalar fields $\{\Xi_I\}_{I=1,\dots,n}$
governed by the action 
\begin{equation}
\label{eq:one}
S = \frac{1}{2} \int \mathrm{d}\tau \, \mathrm{d}^3x 
\left[ \sum_{I=1}^n\partial_\mu\Xi_I \, \partial^\mu\Xi_I 
- \sum_{I,J=1}^n \Xi_I \, \mathbb{M}^2_{IJ} \,\Xi_J \right] \, ,
\end{equation}
where the mass matrix $\mathbb{M}^2$ can be written as
\begin{equation}
\mathbb{M}^2 = \mathbb{C}^T \, \mathbb{M}_d^2\,  \mathbb{C} \, ,
\end{equation}
with $\mathbb{C}$ being an orthogonal matrix and 
$\mathbb{M}_d^2$ a diagonal matrix with eigenvalues
 $\left\{m_1^2,m_2^2\,\cdots\right\}$.
 
One should recall that in a time-dependent background 
the matrices $\mathbb{C}$ and $\mathbb{M}_d^2$ generically depend
on time too. In particular, as mentioned before, 
in an FLRW background these mass eigenvalues include the
gravitational mass squared, $-a''/a$.

As there are many independent quantum fluctuations,
instead of a single wavefunction one is now compelled to consider
an
$n\times n$ matrix, whose entries, $u_{IJ}$,
should be interpreted as the amplitudes of the $I$-th mode
resulting from initial conditions $u_{IJ}\sim \delta_{IJ}\frac{e^{-\imath k\tau}}{\sqrt{2k}}$
for $\tau\to -\infty$.
The Bogolyubov coefficients $\alpha$ and $\beta$ also become matrices, 
which obey the following differential equations:
\begin{equation}
	  \left\{
		\begin{array}{l@{\hspace{2mm}}l}
\alpha' =& -\imath \omega\alpha + \dfrac{1}{2}\omega' \omega^{-1}\beta 
-\mathbb{I}\alpha -\mathbb{J}\beta  \\	\\
\beta' =& \phantom{-} \imath \omega\beta + 
\dfrac{1}{2}\omega' \omega^{-1}\alpha -\mathbb{I}\beta -\mathbb{J}\alpha \ .
		\end{array}
		\right. 
		\label{eq:abdot}
	\end{equation}
Here, the matrices $\mathbb{I}$ and $\mathbb{J}$ are defined as
\cite{Nilles:2001fg}
\begin{eqnarray}
\label{eq:idef}
\mathbb{I} &=& \frac{1}{2} \left[ \sqrt{\omega}\,  \mathbb{C}^T\, \mathbb{C}' \,\sqrt\omega^{-1} + \sqrt{\omega}^{-1} \mathbb{C}^T\, \mathbb{C}' \,\sqrt\omega \right] \\
\nonumber\\
\label{eq:jdef}
\mathbb{J} &=& \frac{1}{2} \left[ \sqrt{\omega} \,\mathbb{C}^T\, \mathbb{C}' \sqrt\omega^{-1} - \sqrt{\omega}^{-1} \mathbb{C}^T\, \mathbb{C}' \sqrt\omega \right]  \, ,
\end{eqnarray}
in which the diagonal matrix $\omega$ is given by
\begin{equation}
\omega^2 = k^2 \mathbf{1} +\mathbb{M}_d^2 \, .
\end{equation}

Solving Eqs.~\eqref{eq:abdot}  
is, in general, rather complicated.
In the remainder of this analysis, we shall simplify the discussion in three ways. 
First, we will only consider two-field inflation models.
In most cases, models 
with two fields
exhibit all the essential features of 
more general multiple-field inflationary models, 
but are much simpler to analyse.
Our second assumption deals with the shape of the potential. 
We  
assume  
that before and after the turn
the eigenvalues of the mass matrix $\mathbb{M}_d^2$ are $m^2$ and $M^2$, 
and that at these times $\frac{\d \mathbb{C}}{\d \tau}=0$.
For successful inflation, the smaller of these eigenvalues, $m^2$, should be much smaller than $H^2$ and thus
we take it to be
negligible in our analysis. 
Finally, we assume that the turn is fast, i.e.,~that $|\dot\vartheta|\ll H$ except for a fraction of an e-fold. This  
will allow us to
neglect the expansion of the universe for the duration of the turn and perform the calculations in Minkowski spacetime,
not having to distinguish between the cosmic time $t$ and the conformal time $\tau$.

With these assumptions, the matrices $\mathbb{I}$ and $\mathbb{J}$ 
defined in Eqs. (\ref{eq:idef}) and (\ref{eq:jdef})
become very simple
\begin{eqnarray}
\label{eq:isim}
\mathbb{I}&=& \frac{1}{2}\dot\vartheta\left(\sqrt{r}+\frac{1}{\sqrt{r}}\right) \left( \begin{array}{cc} 0 & -1 \\ 1 & 0 \end{array}\right) \\ \nonumber\\
\label{eq:jsim}
\mathbb{J}&=& \frac{1}{2}\dot\vartheta\left(\sqrt{r}-\frac{1}{\sqrt{r}}\right) \left( \begin{array}{cc} 0 & 1 \\ 1 & 0 \end{array}\right) \, , 
\end{eqnarray}
where $r=\sqrt{(k^2+M^2)/(k^2+m^2)}>1$ is the ratio of the eigenvalues of $\omega$.
In the fast turn approximation, $|\dot\vartheta|\gg \left\{k,M,H,m\right\}$ 
(that is, $\dot\vartheta$ is the largest scale in the 
problem)\footnote{For simplicity, we assume that the eigenvalues of 
                    $\omega^2$ remain practically constant throughout 
                    the short time of the turn, 
                    as forces inducing the turn 
                    only require a transient presence of 
                    a first derivative of the 
                    potential in a specific direction in the field space. 
                    While in practice the deviations from 
                    this assumption can be sizeable, 
                    the relative change of the eigenvalues 
                    can be much smaller 
                    than $\dot\vartheta$, which justifies neglecting the 
                    first two terms on the right-hand sides 
                    of Eqs.~\eqref{eq:abdot}.},
Eqs.~\eqref{eq:abdot} can therefore be written as
\begin{equation}
\frac{\mathrm{d}\phantom{t}}{\mathrm{dt}} (\alpha\pm\beta) = -(\mathbb{I}\pm \mathbb{J})(\alpha\pm\beta) \, .
\end{equation}
Their solution 
takes the form
\begin{equation}
\label{eq:sol}
(\alpha\pm\beta) = \mathrm{exp}\left( -\int^t (\mathbb{I}(t')\pm \mathbb{J}(t')) \mathrm{d}t' \right)(\alpha\pm\beta)_0 \, ,
\end{equation}
where the subscript `$0$' 
refers to the initial conditions $\alpha_0=\mathbf{1}$, $\beta_0=0$.

Plugging Eqs. (\ref{eq:isim}) and (\ref{eq:jsim}) into
 Eq. (\ref{eq:sol}), we find under these assumptions
\begin{eqnarray}
\label{eq:alphasol}
\alpha &=& \cos\Delta\vartheta \,\left( \begin{array}{cc} 1 & 0 \\ 0 & 1 \end{array}\right)  + \frac{1}{2}\left( \sqrt{r}+\frac{1}{\sqrt{r}}\right)\sin\Delta\vartheta \,\left( \begin{array}{cc} 0 & 1 \\ -1 & 0 \end{array}\right)\ , \ \ \textrm{and} \\
\beta &=& -\frac{1}{2}\left( \sqrt{r}-\frac{1}{\sqrt{r}}\right)\sin\Delta\vartheta\,\left( \begin{array}{cc} 0 & 1 \\ 1 & 0 \end{array}\right) \, ,
\end{eqnarray}
where $\Delta\vartheta$ is the angle by which the inflationary trajectory turns.
This means that both modes,  
light and 
heavy, can be produced at the turn with equal occupation numbers
\be
\label{eq:occan}
|\beta|^2=\frac{1}{4}(\sqrt{r}-1/\sqrt{r})^2\sin^2\Delta\vartheta \, .
\ee
We note in passing that for $r\to 1$, which corresponds 
to the sub-mass-horizon
limit $k\gg \left\{M,H,m\right\}$,
we have $|\beta|^2\simeq \frac{1}{16} \left(\frac{M}{k}\right)^4$, while $\alpha$ becomes a pure
rotation matrix, in agreement with our discussion in \S \ref{sec:f1a}. Provided that $\omega^2$ is positively defined, 
the matrices $\alpha$ and $\beta$ obey the relation $\alpha\alpha^\dagger-\beta\beta^\dagger=\mathbf{1}$,
as the coefficients of the Bogolyubov transformation should.

Finally, we can estimate the energy density in non-relativistic 
heavy particles produced at the turn, as
\be
\label{eq:eden}
\rho_h = \int \frac{\mathrm{d}^3k}{(2\pi)^3} \sqrt{k^2+M^2}\, 
|\beta|^2 \approx \frac{M^4}{16\pi^2}\sin^2\Delta\vartheta \, .
\ee 
To evaluate the integral in Eq. (\ref{eq:eden}), 
we have only included the modes for which $H<k<M$, 
used the expansion $|\beta|^2\sim \frac{M}{k}\sin^2\Delta\vartheta$
valid for $k\ll M$ and neglected 
$\mathcal{O}(H^2/M^2)$ terms.\footnote{The integral can be
                                      calculated without expanding 
                                      $|\beta|^2$, but the rather 
                                      complicated final result differs
                                      from the estimate above 
                                      by only $\sim 10\%$.
                                      Therefore, we take the result 
                                      \eqref{eq:occan} to provide
                                      a sufficient estimate of this effect.}
The obtained energy density scales as $M^4$, which is
\emph{a priori} unrelated to the potential energy density 
during inflation, $V\sim H^2M_P^2$. 
With sufficiently large $M$, one can envision that the entire kinetic density of
the fields corresponding to the velocity component orthogonal to that of the inflaton (eigenvector of $\mathbb{M}$ with
the smallest eigenvalue) is 
utilised for particle production at the turn. 
This is consistent with the scenario in which at the vicinity of the 
turn the slope of the potential is essentially negligible, and 
one can consider it to be flat to a good approximation.   

\subsection{Numerical analysis}
\label{ssnum}

In order to corroborate our analytical results, we will now calculate
$|\beta|^2$ numerically
in the model  
discussed in \S \ref{sec:f1b}.
This can be achieved by  solving the equations for
the wavefunctions of the perturbations, Eqs. (\ref{eq:glmeom1a}) and (\ref{eq:glmeom1b}),
and then employing a generalization of Eq.~(\ref{eq:bet1}) to read off $\beta$ from the wavefunction.
The relation in question is 
\be
\label{eq:bet2}
\beta = \frac{1}{\sqrt{2}}\sqrt{\omega} \,u-\frac{\imath}{\sqrt{2}}\sqrt{\omega}{}^{-1}\left[ u'-\vartheta'\,  \left( \begin{array}{cc} 0 & 1 \\ -1 & 0  \end{array} \right)\, u \right] \, ,
\ee
where $u$ is now to be understood as a $2\times2$ matrix
with the following components
\be
u = \left( \begin{array}{cc} u_l^\mathrm{initial} & u_l^\mathrm{induced} \\ u_h^\mathrm{initial} & u_h^\mathrm{induced}  \end{array} \right) \, .
\ee
As before, the subscripts ${l,h}$ refer to the light 
and heavy components, 
respectively, and the
definition of the initial and induced components
is 
the same as in \S \ref{sec:f1b}.
Values of $\beta$ were determined on an interval of $0.2$ e-folds 
beginning $0.1$ e-folds after the turn (when the rapid
growth of $\beta$ has ceased); as the heavy mode interacts with 
the light one at the turn, its evolution is more complicated
than that of a massive mode in the de Sitter spacetime, 
which leads to small oscillations in $\beta$.
Hence, sampling is necessary
for estimating the accuracy of the determined $|\beta|^2$.

\begin{table}

	\tablepreamble
		\setlength{\tabcolsep}{15pt}
	\sbox{\tableA}{%
		\begin{tabular}{qQ|qQqQ|q}
			 	\toprule			
			\multicolumn{2}{c}{\textbf{model parameters}}
			& \multicolumn{4}{c}{\textbf{derived parameters}}
			& \multicolumn{1}{c}{\textbf{label}}
			\\

\multicolumn{1}{c}{$M/\mpl$}
		 	&
		 	\multicolumn{1}{c|}{$s\mpl$} &
		 	\multicolumn{1}{c}{$M/H$} &
		 	\multicolumn{1}{c}{$\Omega$} &
		 	\multicolumn{1}{c}{$\mathrm{max}\left(\frac{\dot\vartheta}{H}\right)$} &
		 	\multicolumn{1}{c|}{width of $\frac{\dot\vartheta}{H}$} &
		 	\multicolumn{1}{c}{$\phantom{a}$}
		 	
		 	\\

			\midrule

			10^{-4} & 5\sqrt{3}\times 10^3 & 10 & 0.63 & 13 & 0.018 & \textsl{10} 
			\\

			\cmidrule{1-7}

			2\times 10^{-4} & \sqrt{3}\times 10^3 & 20 & 0.31 & 2.8 & 1.0 & \textsl{20 slow}
			\\


			2\times 10^{-4} & 5\sqrt{3}\times 10^3 & 20 & 0.31 & 14 & 0.018 & \textsl{20}
			\\

2\times 10^{-4} & \sqrt{3}\times 10^4 & 20 & 0.31 & 27 & 0.010 & \textsl{20 fast}
			\\
			\cmidrule{1-7}
4\times 10^{-4} & 5\sqrt{3}\times 10^3 & 40 & 0.155 & 14 & 0.019 & \textsl{40}\\

			\bottomrule

		\end{tabular}
	}
	\settowidth{\tblw}{\usebox{\tableA}}
	\addtolength{\tblw}{-1em}

	\begin{center}
		\usebox{\tableA}
	\end{center}

	\renewcommand{\arraystretch}{1}

	\caption{\sl Parameter choices for the model described in \S \ref{sec:f1b} used in the numerical study and with the model labels used in 
Figure \ref{f3}.
	\label{table1}}
	\end{table}

In order to study how the results for $|\beta|^2$ depend on various parameters, we studied five different parameter
choices shown in Table \ref{table1}. We chose three different values of the mass of the heavy field, satisfying
$\frac{M}{H}=10,\,20\,\,\textrm{and}\,\,40$. In the absence of particle production and with the direction of the inflationary trajectory
parametrised by an angle $\theta$, the function $\sin\theta$
expressed as a function of number of e-folds would exhibit damped harmonic oscillations, i.e.,~behave as $\sin(\Omega N+\psi_0)$---the dimensionless frequency
of these oscillations, $\Omega$, is shown in Table \ref{table1}.
The time dependence of the rotation angle $\vartheta$ diagonalizing
the mass matrix is such that $\frac{\dot\vartheta}{H}$ is a Gaussian-like function (but with noticeable deviations from the exact Gaussian 
shape). In Table \ref{table1}, we give the maximal value of this function and the width of this function at half maximum in the number of e-folds.

\begin{figure}
\begin{center}
\includegraphics*[width=10cm]{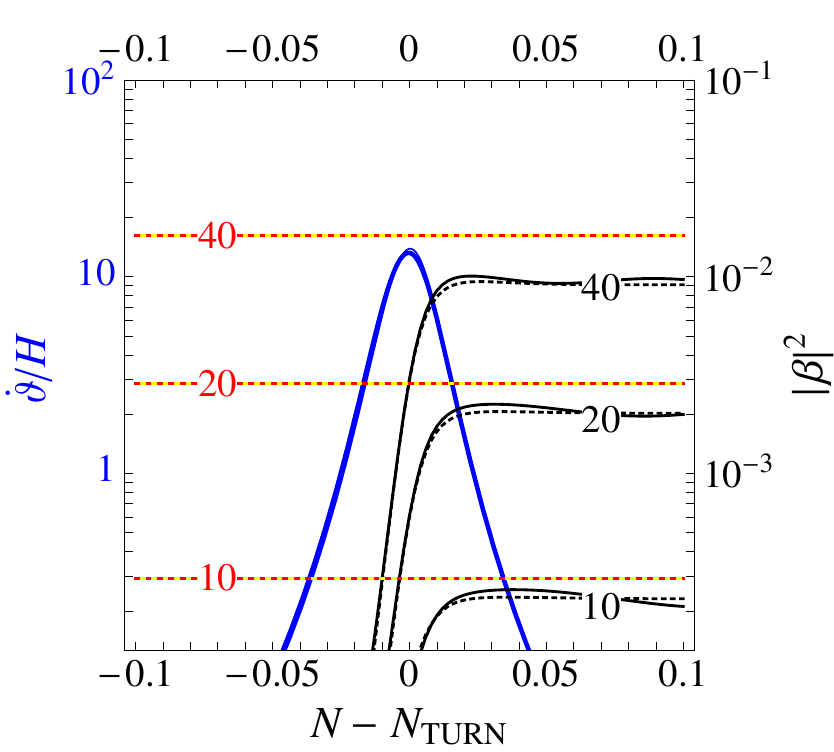}
\end{center}
\caption{\sl Evolution of the occupation number of the light and the heavy mode (dashed and solid black lines, respectively)
for several models listed in Table \ref{table1}. 
For reference, we also show the evolution
of $\frac{\dot\vartheta}{H}$ (blue line) and the dashed red line corresponds to the analytical result \eqref{eq:occan}
for the occupation number.
 \label{f3a}}
\end{figure}

In Figure \ref{f3a}, we show the time evolution of the occupation number for the light and the heavy mode leaving
the Hubble radius 3 e-folds after the turn. The buildup
of the occupation number, which is very similar for the light and the heavy mode, 
takes place during the time at which the eigenvector of the mass matrix are fast changing
(of the two eigenvalues of the mass matrix, the larger one is practically constant and the smaller one is always much
smaller than $H$). The analytical result \eqref{eq:occan} is in qualitative agreement with the numerical solutions.
Note that for the discussed modes our analytical result is on the verge of applicability, as the turn happens close to
the mass radius crossing, $k=Ma$, of the heavy mode.

\begin{figure}
\begin{center}
\includegraphics*[width=10cm]{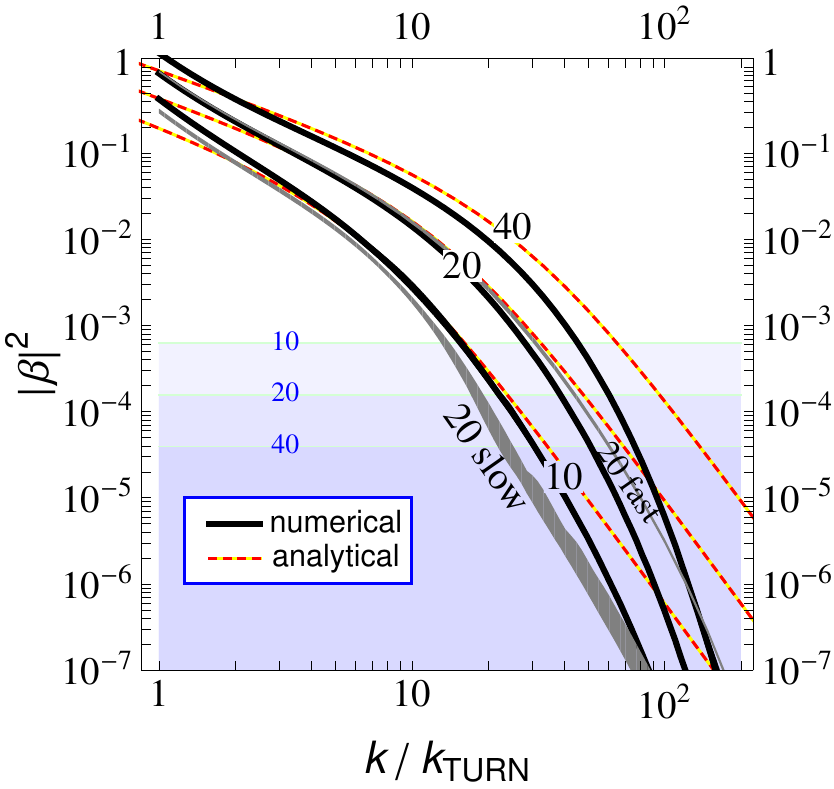}
\end{center}
\caption{\sl Occupation numbers $|\beta|^2$ for different values of $k/k_\mathrm{turn}$,
where $k_\mathrm{turn}$ is the comoving scale that exits the Hubble radius at the turn,
and for three values $M/H=10,\,20,\,40$. Solid lines correspond to the full numerical
calculation outlined in \S \ref{ssnum}, while the dashed lines correspond to
the analytical estimate \eqref{eq:occan}.
The shaded areas correspond to the de Sitter result $|\beta_\mathrm{dS}|^2=\frac{H^2}{16M^2}$,
which is much smaller than $|\beta|^2$
in a wide range of $k$. 
 \label{f3}}
\end{figure}

In Figure \ref{f3}, we show the results for the occupation number of the heavy mode as a function
of the comoving wavenumber $k$. 
Our analytical result agrees with the outcome of our numerical calculation within its range applicability:
deviations are seen when the turn takes place after the mass horizon crossing of the heavy mode (i.e.,~for $M/H=40$)
or the turning rate is not the largest scale. The latter is visible in the behavior of $|\beta|^2$ in the large $k$
limit and for the mode with a reduced turn speed labeled {\sl 20 slow}. Increasing the turn rate (model {\sl 20 fast}) improves
the agreement.
For $k/k_\mathrm{turn}\to1$,
where $k_\mathrm{turn}$ is the comoving scale that exits the Hubble radius at the turn,
we obtain a larger occupation number than given in \eqref{eq:occan}; 
this can be attributed to the fact that the mass squared of
the light field grows transiently negative. 
Modes with $k/k_\mathrm{turn}<1$ are already outside the Hubble radius,
so they have already classicalised and they cannot be interpreted as particles. 
The occupation number of the light mode is very similar
to the result for the heavy mode, therefore, we do not plot it.

In this calculation, we have not included the backreaction of the produced particles on the inflationary background;
we shall return to this issue in \S \ref{s:board}. Here we just note that due to backreaction the occupation numbers
obtained above should be regarded as maximal, subject to energy conservation.

\section{
Particle production in an expanding universe}
\label{sec:particles}

At this point we would like to make a short digression and 
review the ambiguities associated with the concept of 
particle production in a time dependent setting 
\cite{Parker:1969au, Fulling:1979ac,
                            MolinaParis:1997jn, MolinaParis:1999ir,
                            Winitzki:2005rw}.
The quest for 
understanding the nature of this phenomenon
 has lasted for several decades now.
When can we interpret $|\beta_k|^2$ as a measure 
of particle production?

This question is 
closely related with the concept of classicalisation
of quantum perturbations in cosmology. It
can be briefly phrased as follows. 
Deep inside the Hubble radius, the fluctuations
are fundamentally quantum-mechanical and the solutions
to the equations of motion for the field perturbations can be described by 
a linear combination of creation and annihilation operators. These operators 
span the particle Hilbert space in which the vacuum is, by definition, 
annihilated by the destruction operator. 
A \emph{few} e-folds after exiting the horizon, the scalar perturbations 
are said to \emph{classicalise}, in the traditional sense that the position and 
momentum of the wavepacket 
of quantum fluctuations
can be measured simultaneously. 
Said differently, the quantum uncertainty 
inherent to the scalar field becomes negligible.

Where such transition precisely occurs is still a 
matter of dispute. 
Recent work in the literature
has focused on the location of the 
matching surface between quantum and classical
eras \cite{Polarski:1994rz, Polarski:1995jg, 
                            Leach:2000yw,Nalson:2011gc, 
                            Dias:2012qy,Nalson:2013tm}. 
This is undeniably of theoretical
interest but 
perhaps even more importantly can induce an error
in the precision with which one predicts
the value of imprint of the curvature perturbation
in the CMB
radiation (see Ref. \cite{Avgoustidis:2011em},
for example). Another approach to this question
is to replace the discussion of the quantum to 
classical transition with a transition from 
hard to soft subprocesses using the methods
 of the 
renormalization group, 
which can be 
properly adapted to 
time-varying backgrounds \cite{Dias:2012qy}.

Whatever the origin, 
there is little doubt that this classicalisation 
process ought to occur \emph{some} e-folds after
horizon crossing by the time when the decaying mode (in the context of 
inflation) has become negligible. Then, the perturbation is solely 
characterised by the surviving growing contribution, which for 
single-clock inflation can be related to the constant mode associated 
with the primordial fluctuation.
In this regime, it is said the WKB classicality condition
has been reached \cite{Lyth:2006qz}, 
and the commutation relations between two fields
ascribed different momenta vanish.

\subsection{Decoherence effects and particle production}
The concern of where in the e-folding history the transition
from quantum to classical regimes takes place is 
also sometimes
phrased in terms of the decoherence phenomenon, after which one can 
refer to the particle state as a classical condensate \cite{Sakagami:1987mp}.
Mathematically, 
we say that decoherence has occurred if the density matrix
(which also includes the interaction Lagrangian in addition to the free
action)  
is in the diagonal form. The origin of this process
is, however, far from being understood. 
Nevertheless, once decoherence (or classicalisation) has occurred, 
the classical evolution of the perturbations is then determined 
by the separate universe approach (also known as the 
$\delta N$ technique) \cite{Lyth:1984gv, Starobinsky:1986fxa, 
                                Sasaki:1995aw,Sasaki:1998ug, 
                                Lyth:2003im, Rigopoulos:2003ak, 
                                Lyth:2005fi, Seery:2012vj, Mulryne:2013uka}.
Strictly, this evolution is uniquely determined on super-horizon scales
when the perturbations are smoothed over a scale much bigger than the
size of the horizon, $H^{-1}$, but it is independent of the number of 
fields which contribute (or not) to the primordial 
curvature fluctuation.\footnote{Recently, Mulryne 
                \cite{Mulryne:2013uka} has extended
                this formalism to sub-horizon scales, 
                allowing for the quantum statistical 
                properties of the field fluctuations to be 
                described at all scales by a unique set of 
                differential equations.}

In addition to this quantum to classical transition ambiguity, 
the concept of particle itself is also unclear and in some sense
related to this first issue. There are two reasons which make 
the discussion around particle production convoluted. On the one hand, 
to describe particle production one needs well-defined initial conditions
which describe the vacuum 
state of the background \cite{MolinaParis:1997jn, MolinaParis:1999ir}. 
A time-dependent background
will, however, violate in some region in parameter space the adiabaticity 
condition,  
which for modes $k$ with frequency $\omega$ reads
\begin{equation*}
\dfrac{|\dot{\omega}_k|}{\omega_k^2} \ll 1\ \ .
\end{equation*}
If the condition above is not verified, then particle production 
occurs \cite{Birrell:1982ix}.
Consequently, the vacuum of the theory will have to be 
 systematically adjusted. However, a more 
 crucial problem arises as a direct result of
the time-dependent background, 
 which explicitly breaks Poincar\'{e} invariance---inertial 
 observers do
 not agree on each other's definitions of vacuum. 
 This adds to the problematic interpretation of 
 particle production, when compared to a clearer
 result in Minkowski spacetime.\footnote{A possible 
                             resolution to this issue is to 
                             recall that in the infinite past, when 
                             the modes are deep inside the horizon, 
                             spacetime is effectively 
                             Minkowski, and there all
                             inertial observers agree with the 
                             same vacuum state.}
On the other hand, in such a setting the Heisenberg uncertainty relations
prompt to an uncertainty in the particle number owing to 
the possibility of creating virtual pairs of particle--anti-particle.

Yet another factor to take into account is the detection 
of the produced particles, which can 
depend on the level of the coarsening of measurements themselves. 
This is usually a concern of 
the particle-wave duality, which is again related with the classicalisation 
process. As explained  
by Birrell \& Davies \cite{Birrell:1982ix}, the 
detection of a particle is only unambiguous in static 
spacetimes---even in the case of asymptotically static spacetimes 
it is  
not clear 
how to define the volume in momentum space required 
to compute a meaningful (expectation value of the) particle number
density.

For these reasons it is unclear how to define a particle
or confirm the detection of a particle in curved spacetime. The discussion of the 
previous sections, however, evades this issue since by studying
particle production related to sub-horizon scales 
owing to a sharp turn in field configuration,
thereby ignoring the local curvature and taking spacetime
as effectively Minkowski (in the vicinity of the turn).
Moreover, if there were any effects on super-horizon scales, 
these would only add to the calculation presented before, and as 
such the results of the previous sections can be regarded as 
a conservative estimate of particle production.

\subsection{Effects of the turn on observables}
\label{s:board}

Having presented our arguments for particle production at a turn of the inflationary
trajectory, we will now proceed to discussing the impact of this phenomenon on
the power spectrum of the curvature perturbations. We first note that the
turn---and the burst of particle production---lasts only for a fraction of an e-fold.
The timescale of this event is therefore much smaller than the Hubble time,
$H^{-1}$, which allows for the application of the energy conservation law,
as the gravitational effects can be neglected on such small timescales.

The energy emitted in the form of the particles is drained from the kinetic
energy of the scalar fields. Immediately after the turn, the energy density of the
universe consists of  
potential energy, 
and the subdominant components
of kinetic energy of the scalar fields and a collection of particles.
As the kinetic energy is reduced by particle production, the field velocity after
the turn is smaller than the one inferred  
from the slow-roll relation,
which was satisfied before the turn.

Particle production therefore changes the composition of the background energy density by introducing heavy particles,
whose energy density redshifts as $a^{-3}$. However, classical oscillations of a heavy scalar field around a minimum
of the potential also redshift as $a^{-3}$, so the energy transfer does not change the time dependence
of the Hubble parameter.\footnote{While the evolution of the {\sl total} energy 
density does not change, the {\sl kinetic} energy of the scalar fields
exhibits different behaviours: during the oscillations of the heavy field, 
the kinetic energy is periodically transferred into
potential energy and {\sl vice versa}; this effect is absent if the energy is transferred into particles. 
Such oscillation can be in resonance with the mode functions, see \cite{Chen:2011zf, Gao:2013ota}. 
Owing to the smallness of the slow-roll parameter $\varepsilon$,
the effect is, however,
very small in the models considered here.}
Light modes are produced at the turn, too, with a similar occupation number compared
to the heavy ones. This can further reduce kinetic energy after the turn, transferring it into
component that redshifts as $a^{-4}$, but these changes in kinetic energy are
also unimportant for small values of the slow-roll parameter $\varepsilon$.

There is, however, one effect which can have 
a non-trivial impact on the predictions for
the power spectra of the perturbations. 
As it is clear from Eqs.~\eqref{eq:glmeom1a}--\eqref{eq:glmeom1b},
the coupling between the light and the heavy modes depends on the time evolution
of $\vartheta$, the rotation angle that diagonalizes the mass matrix of the 
perturbations.\footnote{For modes with very large $k$, the actual direction of the field velocity or changes in this coupling are 
to a large extent irrelevant---as 
we argued in \S \ref{sec:f1b},  
the oscillations of the
inflationary trajectory around the flat valley of the potential
do not affect
the power spectrum for the modes which are well within their mass
horizon, $k=aM$, at the turn.}
A sudden decrease of the kinetic energy by particle production affects $\vartheta$
in two ways: 
a sudden change in the field velocity results in a jump of $\vartheta'$ and a delta-like
singularity in $\vartheta''$. 
The slowing down of the fields also 
increases the amount
of time in which the fields spend at the turn of the inflationary potential, thereby
lowering $\vartheta'$ after particle production with respect to the situation
in which particle production is neglected. 


\begin{table}

	\tablepreamble
		\setlength{\tabcolsep}{15pt}
	\sbox{\tableA}{%
		\begin{tabular}{qQ|qQ|q}
			 	\toprule			
			\multicolumn{2}{c}{\textbf{model parameters}}
			& \multicolumn{2}{c}{\textbf{derived parameters}}
			& \multicolumn{1}{c}{\textbf{label}}
			\\

\multicolumn{1}{c}{$M/\mpl$}
		 	&
		 	\multicolumn{1}{c|}{$s\mpl$} &
		 	\multicolumn{1}{c}{$M/H$} &
		 	\multicolumn{1}{c|}{$\mathrm{max}\left(\frac{\dot\vartheta}{H}\right)$} & \multicolumn{1}{c}{$\phantom{a}$}
		 	\\
		 	\midrule
			5\times 10^{-4} & 5\sqrt{3}\times 10^3 & 50 & 14 & \textsl{50}\\
10^{-3} & 5\sqrt{3}\times 10^3 & 100 & 14 & \textsl{100}\\
			2\times 10^{-3} & 5\sqrt{3}\times 10^3 & 200 & 14 & \textsl{200} \\
				\cmidrule{1-5}
		
2\times 10^{-3} & \frac{5\sqrt{3}}{2}\times 10^3 & 200 & 7 & \textsl{200 slow} \\
			\bottomrule

		\end{tabular}
	}
	\settowidth{\tblw}{\usebox{\tableA}}
	\addtolength{\tblw}{-1em}

	\begin{center}
		\usebox{\tableA}
	\end{center}

	\renewcommand{\arraystretch}{1}

	\caption{\sl Parameter choices for the model described in \S \ref{s:board} used in the numerical study and with the model labels used in 
Figure \ref{fps}.
	\label{t:s4}}
	\end{table}

\begin{figure}
\begin{center}
\includegraphics*[width=10cm]{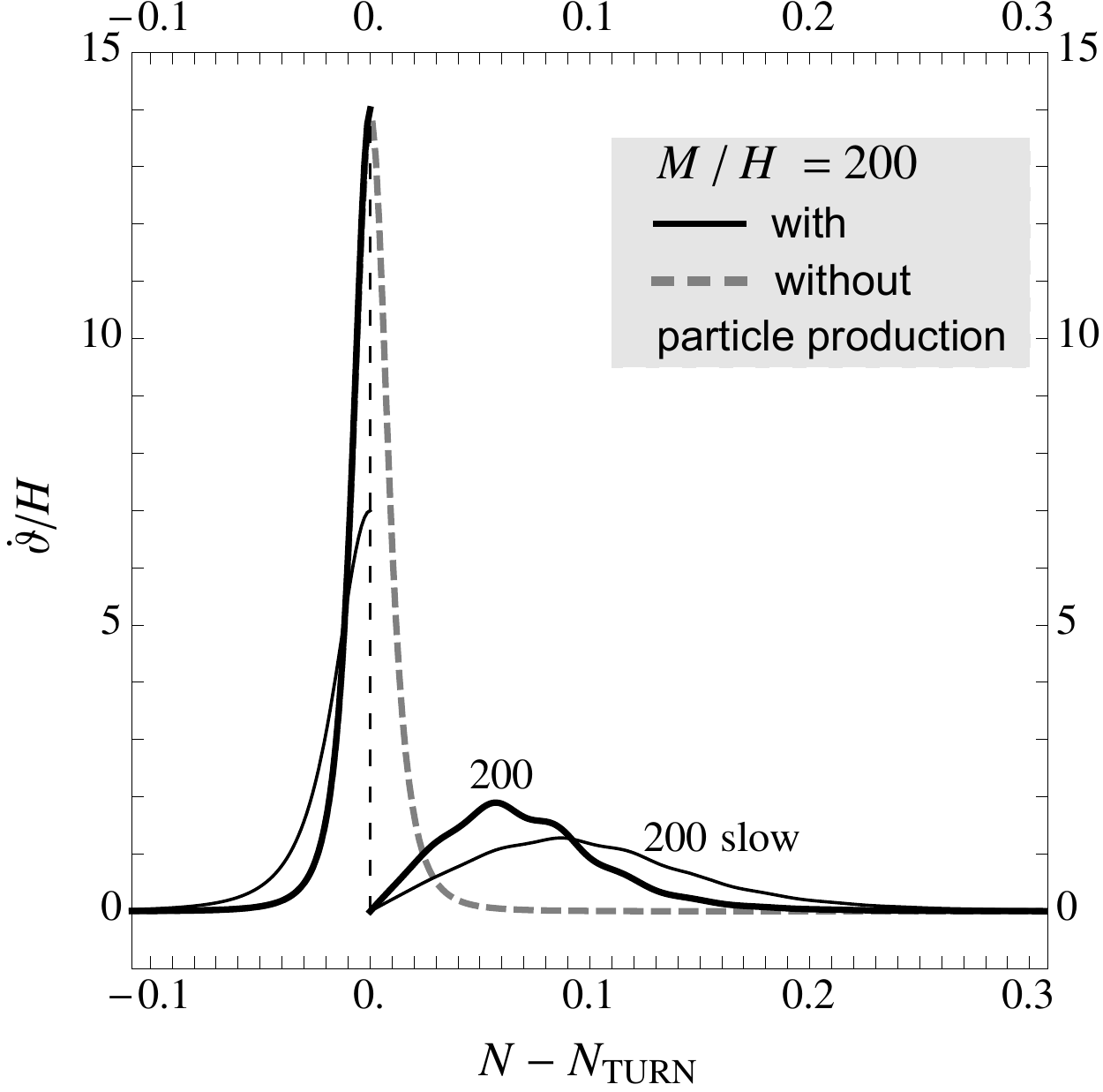}
\end{center}
\caption{\sl Evolution of $\dot\vartheta/H$
for models defined in Table \ref{t:s4}, with $M/H=200$. Black sold (gray dashed) lines correspond to the results obtained with (without) taking particle production into account.
 \label{fth}}
\end{figure} 

A direct comparison of the kinetic energy in the model analysed numerically in \S \ref{ssnum} (which is
of order
$\varepsilon H^2 \sim 10^{-14}$) with 
the estimate in Eq. \eqref{eq:eden} of the maximal energy density
in produced particles
(which gives $\rho_h\sim 10^{-17}$ for $M/H=40$) shows that for the models discussed
in \S \ref{ssnum} the backreaction of the produced particles onto the inflationary background is negligible, in accordance with
the assumption therein.
In order to illustrate the effect of particle production on the power spectrum, we need a model with a smaller
inflaton velocity or a larger mass of the heavy field. Here, we chose that last option and studied numerically
models similar to that  
in Table \ref{table1}, but with a larger $M/H$ equal to 50, 100 and 200, 
assuming that the approximation \eqref{eq:eden}
for the maximal energy density
in produced particles
holds. Then the amount of produced particles is sufficient to reduce the kinetic energy of the fields to zero 
($M/H=200$), to approximately one-third of its initial value ($M/H=100$) or to a few per cent of its initial value ($M/H=50$). 
The parameters of the models are given in Table \ref{t:s4}

Figure \ref{fth} shows how the angle $\vartheta$ changes with and without particle production for the case $M/H=200$. There is a clear difference
between the two cases with $\vartheta$ changing more slowly after kinetic energy of the fields is transferred into particles.
In this case, we can describe the evolution of ${\vartheta'}$ across the turn as
\be
\label{eq:varthp}
\vartheta'(\tau) = \Delta\vartheta'\,\Theta(\tau-\tau_\mathrm{turn})\, ,
\ee
where $\Theta$ is the step function and $\Delta\vartheta'$ is the magnitude of the jump of ${\vartheta'}$.
Solving the equations of motion \eqref{eq:glmeom1a}--\eqref{eq:glmeom1b} for the perturbations, we should
therefore implement the following matching relations
\be
\label{eq:match}
v'_l(\tau_+) = v'_l(\tau_-) +\Delta\vartheta'\,v_h(\tau_-)\,,\qquad v'_h(\tau_+) = v'_l(\tau_-) -\Delta\vartheta'\,v_l(\tau_-) \, , 
\ee
where, as in \S\ref{sec:f1a}, an infinitesimally small interval
$(\tau_-,\tau_+)$ contains $\tau_\mathrm{turn}$.
The results for $M/H$ are almost indistinguishable from those shown in Figure \ref{fth}; the only difference is
that the small oscillatory features of $\frac{\dot\vartheta}{H}$ after particle production are less pronounced.

\begin{figure}
\begin{center}
\includegraphics*[width=10cm]{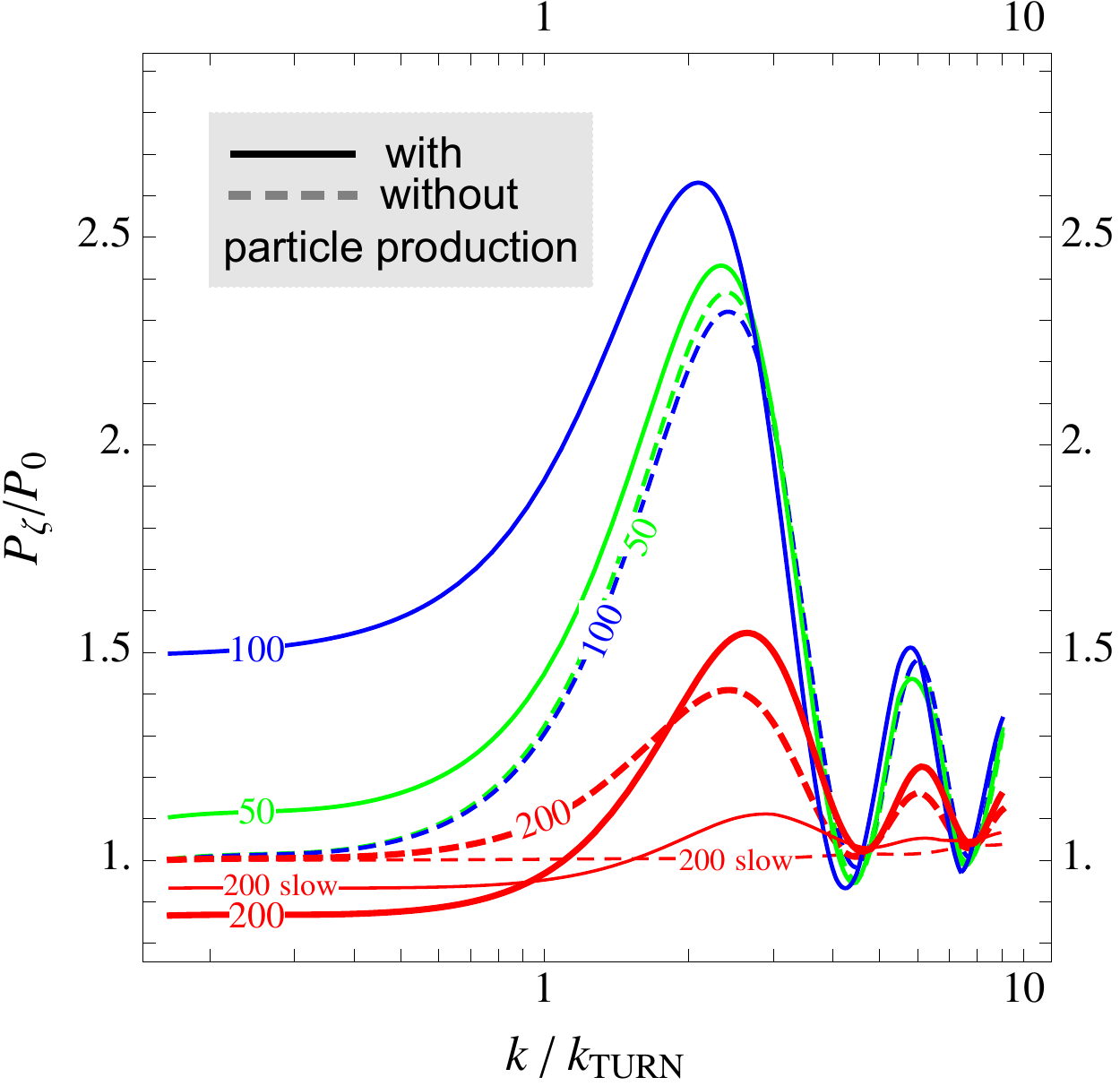}
\end{center}
\caption{\sl Power spectrum of curvature perturbations
for the models given in Table \ref{t:s4}, with $M/H=50$, $100$ and $200$, normalized to the single-field prediction (green, blue and red lines,
respectively). Solid (dashed) lines correspond to the results obtained with (without) taking particle production into account.
 \label{fps}}
\end{figure}

Applying the modified time evolution of $\vartheta$ into the equations of motion for the perturbations
leads to a modification of the power spectrum shown in Figure \ref{fps}. 
The most significant modification can be seen for $k<k_\mathrm{turn}$, {\em i.e}., for
the modes which are already super-Hubble at the time of the turn. This can be understood
in the following way, given our description
of particle production. At the beginning of the turn the light mode acts as a source for the heavy mode,
whose initial amplitude has already decayed. At the instant of particle production,
there is a backreaction of this induced heavy mode on the light one, as described
by the matching conditions (\ref{eq:match}). Thus, the light mode receives additional contribution,
whose phase depends on the evolution of the heavy mode between the onset of the turn
and the instant of particle production. This additional contribution can interfere either
constructively or destructively with the initial light mode.

For $k>k_\mathrm{turn}$, the amplitude of the heavy mode is not so small as in the previous case
and the interactions between the modes, which leads to oscillations in the power spectrum
of the curvature perturbations, is not much affected by the altered evolution of $\vartheta$.
Therefore, the magnitude and 
$k$-dependence do not change qualitatively, as should be expected from the discussion above.

\section{Summary}
\label{sec:discuss}

Inflationary phenomenology is an important branch in cosmology
which aims at obtaining observable predictions 
which can probe and determine the nature of the inflationary period.
The current constraints on higher correlation functions
of the primordial fluctuations are compatible with a gaussian distribution \cite{Ade:2013ydc}. 
This implies that non-gaussianity might not be a useful smoking-gun 
for the nature of the microphysics during the early universe. 
In particular, it raises the question of whether we might be able 
to ever break the degeneracy between inflation driven by a single scalar field, 
or multiple degrees of freedom. 

When embedding an inflationary realisation within
a high-energy theory demands dealing with a number of scalar-fields, 
all of which are, in general, active during inflation. 
In this paper we revisit multiple-field inflation models and reexamine the
phenomenology of sharp turns in 
the inflationary trajectory, including  
the advent of particle production
as a result. 
For simplicity we focus on a two-field inflation model and study the 
limiting case when one of the fields is very light, 
whilst the other is very heavy. 
We provide both analytical as well as numerical estimates for the particle numbers
produced as a consequence of the turn. 
We focus on the modes which are sub-horizon when the turn happens, 
to avoid the ambiguities in the concept of particle production after horizon-crossing.

Using the generalised Bogolyubov coefficients, we are able to estimate the
energy of the quanta produced during the turn. If the turn is very sharp, 
then most of the kinetic energy of the fields goes into particle production, 
while the potential energy remains approximately constant. Therefore, this process
increases the amount of time spent by the fields at the turn of the
trajectory and can induce
 a damping in the oscillations of the fields enveloping the inflationary 
flow, as they settle back into the local minima of the potential. Consequently, 
this will result in a modified power spectrum of perturbations.

On the other hand, if we were to ignore the 
production of quanta and focused on sub-horizon modes,  
there  
would be no change in the power spectrum of 
perturbations. 
We conclude that the rich phenomenology of multiple-field
inflation can also lead to a set of parameters which might
be largely unconstrained by observational measurements, which 
was also noted by Elliston \etal~\cite{Elliston:2013afa}.

There are a number of ways in which this analysis can be further 
generalised. Under theoretical motivations, we might wish the 
contemplate the possibility of one or more fields having non-canonical kinetic terms.
This, in 
principle, will change the effective potential between the fields.
One could also imagine 
the effect of dropping the simplifying assumption of 
the initial vacuum state being Bunch--Davies, 
by considering a more generic, mixed state.

Another venue of study could focus on
modes which are super-horizon
when the turn happened.
Consider, for example, double 
quadratic inflation 
\cite{Polarski:1994rz, Silk:1986vc, 
      GarciaBellido:1995qq, Langlois:1999dw, 
      Vernizzi:2006ve}.
In this toy model, 
two canonically normalised 
scalar fields, $\phi$ and $\chi$, which interact through gravity, 
obey the following action
\begin{equation}
S=\frac{1}{2} \int{\d^4 x \, \sqrt{-g}} \ 
\Big\{
R-(\partial_\alpha \phi)^2 -(\partial_\alpha \chi)^2
-m_{\phi}^2 \phi^2-m_{\chi}^2 \chi^2
\Big\}\ .
\end{equation}
It is well known that choosing appropriate initial conditions for the 
values of the fields, there is a turn 
of the inflationary trajectory in field configuration space.
Despite the fluctuations being parallel to the direction of the 
trajectory \emph{before} and \emph{after} the turn, 
resulting in a deflection angle of roughly $\pi/2$, 
the turn itself takes 
a sizeable number of e-folds to conclude. This will be 
the main limiting factor for an appreciable value of $|\beta_k|^2$.
Using our formulae from the previous sections, 
we find that the magnitude of $|\beta_k|^2$ for super-horizon modes
is below the numerical precision, and therefore particle production 
is not sizeable for such modes. Additionally, even if $|\beta_k|^2$
was appreciably large, one would raise the issue of whether one
could interpret this object as a measure of particle 
production since one would expect classicalisation to have occurred.

To conclude, the analysis presented in this paper is a step
forward towards understanding the phenomenology of 
turns in multiple-field inflation and 
their impact on lowest-order statistics. 
Our aim is purely exploratory, 
and we do not attempt to use the \emph{Planck} data
to constrain the parameters of the model we investigated. 
Rather, our analysis intends to illustrate 
as a point of principle the great level of 
complexity involved in a bottom-up approach, by which one 
would be able to learn about the microphysical Lagrangian
from observations of the CMB.

\acknowledgments 
We are grateful to Sera Cremonini and Scott Watson for many discussions, and 
to Anne Davis and Claudia de Rham for useful comments on a draft version of this paper.
Many thanks go to Xian Gao, David Langlois, Shuntaro Mizuno, Toshifumi Noumi and Masahide Yamaguchi
for valuable correspondence.
RHR would like to thank the University of Warsaw and DAMTP at the University of Cambridge for the kind hospitality.
KT would like to acknowledge the hospitality and stimulating
atmospheres of the Lorentz Center and the Case Western Reserve University.
MK and KT are partly supported by grant  IP2011 056971 
from the Ministry of Science and Higher Education.
RHR is supported by the Department of
Energy grant DE-SC0009946.

\section*{Appendix}

Here we collect the formulae for the factors $\mathcal{F}_l$, $\mathcal{F}_h$ and $\mathcal{F}_{lh}$
introduced in Ref.~\cite{Gao:2012uq}. Let $v_l$ and $v_h$ be the wavefunctions of the light and heavy mode, respectively,
in the absence of interaction between the modes, i.e.~for $\vartheta=$const.
One then defines integrals:
\begin{eqnarray}
I_l &=& \mathrm{i} \int\mathrm{d}\tau \left(v_l(\tau)+v_l^\ast(\tau)\right)\vartheta'^2(\tau) v_l(\tau) \nonumber \\
I_h &=& \mathrm{i} \int\mathrm{d}\tau \left(v_l(\tau)+v_l^\ast(\tau)\right)\left( \vartheta''(\tau) v_h(\tau) + 2 \vartheta'(\tau)v'_h(\tau) \right)  \nonumber \\
J_{lh} &=&  \int\mathrm{d}\tau \left(v_l(\tau)+v_l^\ast(\tau)\right)\left( \vartheta''(\tau) v_h(\tau) + 2 \vartheta'(\tau)v'_h(\tau) \right) \int^\tau
\mathrm{d}\tau' v_h^\ast(\tau') \left( \vartheta''(\tau') v_l(\tau') + 2 \vartheta'(\tau')v'_l(\tau') \right) +
\nonumber \\
&&
-\int\mathrm{d}\tau \left(v_l(\tau)+v_l^\ast(\tau)\right)\left( \vartheta''(\tau) v_h^\ast(\tau) + 2 \vartheta'(\tau){v_h^\ast}'(\tau) \right) \int^\tau
\mathrm{d}\tau' v_h(\tau') \left( \vartheta''(\tau') v_l(\tau') + 2 \vartheta'(\tau')v'_l(\tau') \right)
\nonumber 
\end{eqnarray} 
and the correction factors:
$$
\mathcal{F}_l = I_l+I_l^\ast\,,\qquad \mathcal{F}_h=|I_h|^2\,,\qquad \mathcal{F}_{lh} = J_{lh}+J_{lh}^\ast\, .
$$


\bibliographystyle{JHEPmodplain}
\bibliography{references}

\end{document}